\documentclass[preprint]{iopart}
\usepackage{iopams,cite,color}
\usepackage{graphicx}

\def\iu{{\rm i}}

\def\Re{{\rm Re}}
\def\Im{{\rm Im}}

\def\bfe{\boldsymbol E}
\def\bfb{\boldsymbol B}
\def\bfa{\boldsymbol A}
\def\bjs{\boldsymbol{\mathcal J}}
\def\mte{\boldsymbol{\mathcal E}}

\def\bsig{\boldsymbol{\sigma}}

\def\hatp{\widehat {\varphi}}
\def\hatK{\widehat{\mathfrak K}}

\def\ez{\mathbf e_z}

\def\br{\mathbf r}
\def\bx{\bar x}
\def\brx{\breve x}

\def\dx{{\rm d}x}
\def\dzet{{\rm d}\zeta}
\def\dxi{{\rm d}\xi}
\def\dch{{\rm d}\chi}

\def\ke{k_{\rm eff}}
\def\bq{\breve q}
\def\dq{{\rm d}\bq}
\def\ds{{\rm d}s}

\def\sg{{\rm sg}}

\def\Int{\int\!\!\!\int}

\begin{document}

\title{Nonretarded edge plasmon-polaritons in anisotropic two-dimensional materials}

\author{Dionisios Margetis$^{1}$, Matthias Maier$^{2}$, Tobias Stauber$^{3}$, Tony Low$^{4}$ and Mitchell Luskin$^{5}$}

\address{${}^1$Institute for Physical Science and Technology, and Department of Mathematics, and Center for Scientific Computation and Mathematical Modeling,
University of Maryland, College Park, MD 20742, USA}

\address{${}^2$Department of Mathematics, Texas A\&M University, College Station, TX 77843, USA}

\address{${}^3$Materials Science Factory, Instituto de Ciencia de Materiales de Madrid, CSIC, E-28049 Madrid, Spain}

\address{${}^4$Department of Electrical \& Computer Engineering,
University of Minnesota, Minneapolis, Minnesota 55455, USA}

\address{${}^5$School of Mathematics, University of Minnesota, Minneapolis, MN 55455, USA}

\ead{\mailto{dio@math.umd.edu,} \mailto{maier@math.tamu.edu},  \mailto{tobias.stauber@csic.es}, \mailto{tlow@umn.edu} and \mailto{luskin@umn.edu}}

\begin{abstract}
By an integral equation approach to the time-harmonic classical Maxwell equations, we describe the dispersion in the nonretarded frequency regime of the edge plasmon-polariton (EPP) on a semi-infinite flat sheet.  The sheet has an \emph{arbitrary, physically admissible, tensor valued}
and spatially homogeneous conductivity, \color{black} and serves as a model for a family of two-dimensional conducting materials. We formulate a system of integral equations for the electric field tangential to the sheet in a homogeneous and isotropic ambient medium. We show how this system is simplified via a length scale separation. This view entails  the
quasi-electrostatic approximation, by which the tangential electric field is replaced by the gradient of a scalar potential, $\varphi$. By the Wiener-Hopf method, we solve an integral equation for $\varphi$ in some generality. The EPP dispersion relation comes from the elimination of a divergent limiting Fourier integral for $\varphi$ at the edge. We connect the existence, or lack thereof, of the EPP dispersion relation to the {\em index} for Wiener-Hopf integral equations, an integer of topological character. We indicate that the values of this index may express an asymmetry due to the material anisotropy  in the number of wave modes propagating on the sheet away from the edge \color{black} with respect to the EPP direction of propagation. We discuss extensions such as the setting of two semi-infinite, coplanar sheets.  Our theory forms a generalization of the treatment by Volkov and Mikhailov (1988 {\it Sov.\ Phys.\ JETP} \textbf{67} 1639). \color{black}

\end{abstract}

\vspace{2pc}
\vspace{2pc}

\maketitle

%%%%%%%%%%%%%%%%%%%%%%%%%%%%%%%%%%%%%%%%%%%%%%%%%%%%%%%%%%%%%%%
%%%%%%%%%%%%%%%%%%%%%%%%%%%%%%%%%%%%%%%%%%%%%%%%%%%%%%%%%%%%%%%
\section{Introduction}
\label{sec:Intro}
%%%%%%%%%%%%%%%%%%%%%%%%%%%%%%%%%%%%%%%%%%%%%%%%%%%%%%%%%%%%%%%
%%%%%%%%%%%%%%%%%%%%%%%%%%%%%%%%%%%%%%%%%%%%%%%%%%%%%%%%%%%%%%%

The last decade has witnessed rapid advances in the fabrication of two-dimensional (2D) materials such as graphene and van der Waals heterostructures~\cite{Torres2014-book,Geimetal2013,Novoselovetal2012,CastroNeto2009}. Many of these materials have remarkable electronic and thermal transport properties, and exhibit high mechanical flexibility and strength. These attributes enable a wealth of possibilities for nanoscale applications. In particular, some conducting 2D materials such as doped graphene can sustain short-scale electromagnetic surface waves, called surface plasmon-polaritons (SPPs), at terahertz frequencies~\cite{Pitarke2007,Jablan2013,Bludov2013,Lowetal2017,Stauber2014}. SPPs are collective charge excitations of the electron plasma triggered by incident photons, and have wavelengths well below the diffraction limit which is invaluable in nanophotonics.

The following experimental observations for SPPs deserve attention~\cite{Feietal2015,Gonzalez2014,Feietal2012,Chenetal2012,Crasseeetal2012,Yanetal2012,Taoetal2011,Barnesetal2003}:
(i) SPPs can have wavelengths that are small compared to those of free photons at the same frequency in the  mid-infrared \color{black} spectrum. (ii) SPPs are waves confined to the 2D material.
(iii) Their existence and dispersion can be controlled by the conductivity of the 2D material. (iv) The SPPs in the bulk of the 2D material may be accompanied by other short-scale waves, the edge plasmon-polaritons (EPPs), that propagate along edges~\cite{Feietal2015,Crasseeetal2012,Yanetal2012,Taoetal2011}. The EPP exhibits features that are distinct from those of the SPP in the bulk, e.g., a smaller wavelength and confinement near the material edge. Key properties of these waves are plausibly described via particular solutions of classical Maxwell's equations or their approximations; see, e.g.,~\cite{Lowetal2017,Bludov2013,Hanson2008,Hanson2013-E,Nikitinetal2011,Wangetal2011,MML2017,CohenGoldstein2018}.

In this paper, we tackle the following question: What is the dispersion relation of EPPs on semi-infinite flat sheets with straight edges and {\em general, physically admissible tensor valued} yet spatially homogeneous, frequency dependent conductivities? We solve a boundary value problem for time-harmonic Maxwell's equations in the nonretarded frequency regime, when the propagation constant of the EPP is large in magnitude compared to the wave number in the ambient medium.
We formulate integral equations for the electric field tangential to the sheet, and approximately simplify this system by scale separation in a systematic manner. We obtain a singular integral equation for an emerging scalar potential, $\varphi$. By the Wiener-Hopf method, we explicitly solve this equation. The dispersion relation comes from the requirement that $\varphi$ on the plane of the sheet must be bounded and continuous at the edge.
\medskip 

Our contributions with the present work can be described as follows.

\begin{itemize}

\item In some generality, we formulate a system of integral equations for the electric field tangential to the 2D material, in correspondence to a boundary value problem for Maxwell's equations. This formalism may account for a
tensor valued nonlocal conductivity of a sheet surrounded by a homogeneous isotropic medium.

\item By the separation of two length scales, namely, the wavelength of radiation in free space and the EPP length scale on the 2D material, we reduce the above system to an integral equation for a scalar potential, $\varphi$.

\item By the Wiener-Hopf method~\cite{Krein1962,Masujima-book,LawrieAbrahams2007,MML2017}, we solve the integral equation for $\varphi$ exactly in terms of Fourier integrals when the tensor surface conductivity is spatially homogeneous. The EPP dispersion relation
is obtained by elimination of a divergent limiting Fourier integral for $\varphi$ at the edge~\cite{VolkovMikhailov1985}.

\item We connect the existence of the EPP dispersion relation to the index, $\nu_K$, for Wiener-Hopf equations~\cite{Krein1962}, which is an integer of topological character. 

\item We discuss an interpretation of this $\nu_K$ by relating it  to the number of wave modes, including bulk SPPs, that propagate on the sheet away from the edge. \color{black}

\item By the Mellin transform technique, we derive an asymptotic formula of universal character for the EPP wave number in the long-wavelength limit. 

\item We discuss extensions of our formalism, namely, settings with: (i) an isotropic but nonhomogeneous, in the direction vertical to the sheet, ambient medium, and (ii) two semi-infinite coplanar conducting sheets that are in contact along a line.

\end{itemize}

Our work is broadly motivated by experimental designs in the plasmonics of 2D materials~\cite{Lowetal2017}. The technological goal in this direction is to scale down optical devices towards the nanoscale, which calls for manipulating electromagnetic fields at length scales below the diffraction limit~\cite{Jablan2013}.  An emerging class of 2D materials with in-plane anisotropies prompts the question about the features of their EPPs~\cite{Lowetal2017}. \color{black} We study an aspect of this problem by use of classical electromagnetic wave theory. 
Our work has been partly inspired by the quasi-electrostatic approach in~\cite{VolkovMikhailov1988}  where, in the context of an antisymmetric tensor surface conductivity without dissipation,  it is found that the EPP is not always guaranteed to exist, and may enter into the bulk SPP continuum under some specific conditions. \color{black} Our goal is to introduce an analytical characterization of this transition by use of the index for Wiener-Hopf integral equations~\cite{Krein1962}.  Our results would then establish some general conditions for the existence of the EPP in 2D materials with arbitrary, physically admissible conductivity tensors. \color{black}

EPPs on 2D conducting materials such as graphene are intimately related to edge magnetoplasmons. These are highly oscillatory waves that are observed to propagate along the
boundaries of 2D electron plasmas, often in the presence of a transverse static magnetic field~\cite{Grimes1976,Allen1983,Mast1985,Grodnensky1990}. The dispersion relation of the edge magnetoplasmon propagating along the joint boundary of two semi-infinite, coplanar sheets of electron plasmas has been derived in the case with antisymmetric tensor conductivities~\cite{VolkovMikhailov1988}. The analysis in~\cite{VolkovMikhailov1988} starts with an {\em ad hoc} scalar potential, in the quasi-electrostatic approach, and makes use of the Wiener-Hopf method for the related integral equation.

Here, we generalize the analysis of~\cite{VolkovMikhailov1988} in a fourfold way. First, we show how the integral equation for the scalar potential in the nonretarded frequency regime originates from a two-scale analysis of an integral formalism for Maxwell's equations. This description illustrates from a multiscale view the nature of the quasi-electrostatic approach. Second, we derive the EPP dispersion relation for more general tensor surface conductivities. Third, we point out universal features of the EPP, by showing how certain elements of the conductivity tensor enter the long-wavelength limit of the EPP dispersion relation. Fourth, we connect the existence of this relation to a specific (zero) value of the index $\nu_K$ for Wiener-Hopf integral equations, a winding number in a suitably defined complex plane~\cite{Krein1962}.  We indicate that the (zero or nonzero) $\nu_K$ equals half the difference in the number of wave modes that propagate away from the edge under the change in the EPP direction of propagation. This number accounts for both
exponentially decaying and growing waves in the direction normal to the sheet, including bulk SPPs. For a nonzero $\nu_K$, the EPP dispersion relation ceases to exist.\color{black}

In regard to edge magnetoplasmons in 2D materials, there is a sequence of related works that have hydrodynamic ingredients, e.g.,~\cite{Mast1985,Fetter1985,Wuetal1985,Fetter1986,Fetter1986b,Aleiner1994,Rudin1997,Vaman2014,CohenGoldstein2018}. In these treatments, the 2D Euler or Navier-Stokes equations for the electron fluid system, without or with viscosity, are coupled with the Poisson equation for an electrostatic potential in the three-dimensional (3D) space.
In many, albeit not all, of these works the exact kernel of the integral equation for the potential is replaced by a simplified kernel~\cite{Mast1985,Wuetal1985,Fetter1985,Fetter1986,Fetter1986b}. This \emph{ad hoc} approximation is known to be inadequate for capturing the effect of the long-range Coulomb interaction in the electron system, manifested through a logarithmic term in the long-wavelength limit of the phase velocity for the EPP~\cite{Aleiner1994,CohenGoldstein2018}. In our formalism, we employ a tensor conductivity model for the 2D system in the spirit of~\cite{VolkovMikhailov1985,VolkovMikhailov1988}, thus avoiding the use of a hydrodynamic model. In this context, we retain the exact kernel of the electrostatic interaction when we apply the Wiener-Hopf method~\cite{Krein1962,Masujima-book,LawrieAbrahams2007}. Consequently, our result captures the anticipated 
logarithmic divergence of the phase velocity at sufficiently long wavelengths~\cite{CohenGoldstein2018}.

In this paper, we place emphasis on analytical concepts and methods. Numerical simulations for physical predictions are the main subject of a separate paper~\cite{MSLLM-preprint}.

Our model and analysis have some limitations. For example, retardation effects with a tensor conductivity are neglected. The derivation of error estimates for the leading-order scale separation  lies beyond our present scope. (A derivation of the EPP dispersion relation with retardation in isotropic and homogeneous 2D materials can be found in~\cite{DM-preprint}.) 
Effects of nonlocality in the conductivity, e.g., from the viscous electron flow~\cite{CohenGoldstein2018}, are not touched upon. We also neglect the effect that the boundary condition for the electron flow at the edge has on the surface conductivity~\cite{CohenGoldstein2018}.

The remainder of the paper is organized as follows. In section~\ref{sec:formulation-asympt}, we generally formulate a system of integral equations for the electric field tangential to the sheet, and apply a scale separation that leads to the use of a scalar potential, $\varphi$. Section~\ref{sec:solution} addresses the solution of the integral equation for the potential $\varphi$ by the Wiener-Hopf method for a homogeneous sheet and introduces the related concept of the index. In section~\ref{sec:EPP_dispersion}, we derive the EPP dispersion relation for zero index via the continuity of $\varphi$, and outline a few examples. Section~\ref{sec:low_freq} focuses on the long-wavelength asymptotic expansion of the EPP dispersion relation. In section~\ref{sec:extensions}, we provide two extensions in the spirit of~\cite{VolkovMikhailov1988}. Section~\ref{sec:conclusion} contains a summary and discussion of our results. The appendices provide technical yet non-essential derivations for results of the main text.

\medskip

\noindent {\em Notation and terminology}. $\mathbb{R}$ and $\mathbb{Z}$ are the sets of real numbers and integers, respectively, and $\mathbb{C}$ is the complex plane. $\Re\, w$ ($\Im\, w$) denotes the real (imaginary) part of complex $w$. Boldface symbols denote vectors or matrices. The Hermitian part of the matrix $\boldsymbol M$ is $\frac{1}{2}({\boldsymbol M^*}^T+\boldsymbol M)$ where the asterisk (${}^*$) and $T$ as {\em superscripts} denote complex conjugation and transposition, respectively. $F|_\Sigma$ means that the function $F(\br)$ is evaluated if the position vector $\br$ is in set $\Sigma$ ($\Sigma\subset \mathbb{R}^3$). $f=\mathcal O(g)$ ($f=o(g)$) for scalars $f$ and $g$ means that $|f/g|$ is bounded by a nonzero constant (approaches zero) in a prescribed limit. $f\sim g$ implies $f-g=o(g)$. In $Q_\pm(\xi)$ the $\pm$ {\em subscript} indicates a function that is analytic in the upper ($+$, $\Im\xi>0$) or lower ($-$, $\Im\xi<0$) half $\xi$-plane ($\xi\in\mathbb{C}$). The limit $x\downarrow 0$ ($x\uparrow 0$) indicates that the real $x$ approaches 0 from positive (negative) values. The term ``sheet'' has one of the following meanings: either (i) a material sheet, which has negligible thickness, or (ii) a Riemann sheet, which is a branch of a multivalued function of a complex variable. The terms ``top Riemann sheet'' and ``first Riemann sheet'' are used interchangeably. The $e^{-\iu\omega t}$ time dependence is employed throughout where the angular frequency $\omega$ is positive ($\omega>0$, $\iu^2=-1$).\looseness=-1

%%%%%%%%%%%%%%%%%%%%%%%%%%%%%%%%%%%%%%%%%%%%%%%%%%%%%%%%%%%%%%%
%%%%%%%%%%%%%%%%%%%%%%%%%%%%%%%%%%%%%%%%%%%%%%%%%%%%%%%%%%%%%%%
\section{Formulation: System of integral equations and two-scale separation}
\label{sec:formulation-asympt}
%%%%%%%%%%%%%%%%%%%%%%%%%%%%%%%%%%%%%%%%%%%%%%%%%%%%%%%%%%%%%%%
%%%%%%%%%%%%%%%%%%%%%%%%%%%%%%%%%%%%%%%%%%%%%%%%%%%%%%%%%%%%%%%

In this section, we generally formulate a system of integral equations for the electric field tangential
to the plane of a conducting sheet. We apply separation of two length scales, a `fast' and a `slow' one, to reduce this system to a single integral equation.
% for an emerging scalar potential, $\varphi$, on the sheet.

%%%%%%%%%%%%%%%%%FIGURE:Geometry%%%%%%%%%%%%%%%%%%%%%%%%%%%%
\begin{figure}
\begin{center}
\includegraphics*[scale=0.4,trim=-1.5in 2.3in 0in 0.3in]{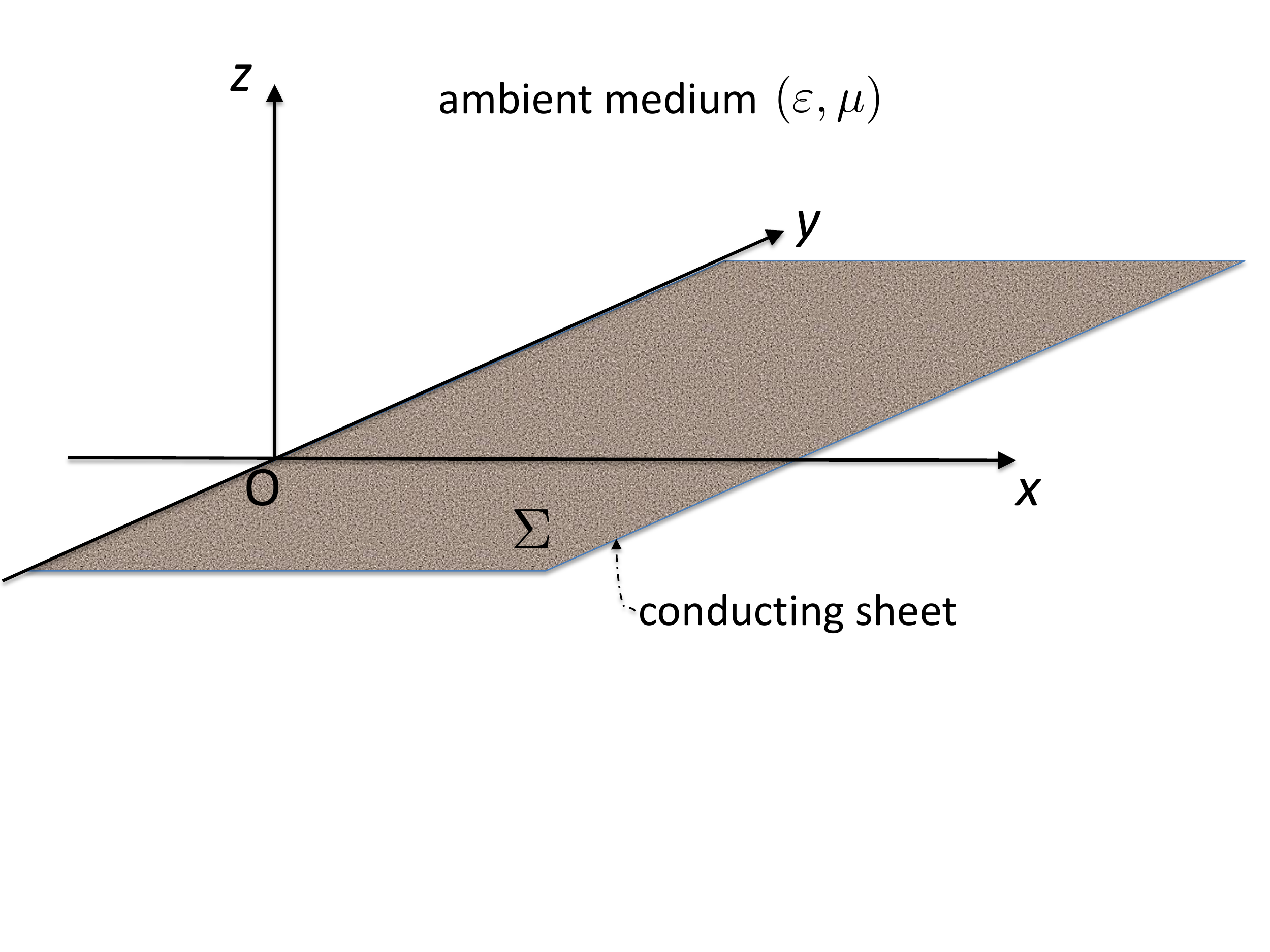}
    \caption{(Color online) Schematic of the geometry. The semi-infinite conducting sheet $\Sigma$ lies in the $xy$-plane, $x>0$. The sheet is immersed
    in a homogeneous and isotropic medium with dielectric permittivity $\varepsilon$ and magnetic permeability $\mu$.}\label{fig:geom}
    \end{center}
\end{figure}
%%%%%%%%%%%%%%%%%%%%%%%%%%%%%%%%%%%%%%%%%%%%%

The geometry is shown in figure~\ref{fig:geom}. Consider a semi-infinite conducting sheet $\Sigma$ that
lies in the $xy$-plane with $x>0$. The ambient space is filled with a homogeneous and isotropic, possibly lossy, medium of dielectric permittivity $\varepsilon$ and magnetic permeability $\mu$. (See section~\ref{sec:extensions} for extensions of this setting).

At this stage, we assume that the electric surface current density, defined on $\Sigma$ (at $z=0$ with $x>0$), is formally described by a vector-valued linear functional,
$\bjs[\bfe_\parallel|_\Sigma]$. The, in principle $\omega$-dependent, functional
$\bjs[\cdot]$ maps a vector field tangential to $\Sigma$ to another vector field tangential to $\Sigma$, and $\bfe_\parallel$ is the electric field parallel to the $xy$-plane. Thus, $\ez\cdot \bjs[\bfe_\parallel|_\Sigma]=0$ where $\mathbf e_\ell$ denotes a unit Cartesian vector ($\ell=x,y,z$).  For example, a nonlocal model for the surface current density in a spatially homogeneous 2D material is described by \color{black}
%%%%%%%%%%%%%%%%%%%%%%
\begin{equation}\label{eq:sigma-nonlocal}
\bjs[\bfe_\parallel|_\Sigma](\br)=\Int\limits_{\Sigma} \check{\bsig}(\br-\br')\cdot\bfe_\parallel(\br')\,{\rm d}\br' \quad (\br\in\Sigma),
\end{equation}
where $\check{\bsig}(\br)$ is a matrix-valued conductivity kernel on $\Sigma$, viz.,
\begin{equation}\label{eq:sigma-tensor}
\check{\bsig}(x,y)=\left(\begin{array}{ccc}
\check\sigma_{xx}(x,y) & \check\sigma_{xy}(x,y) & 0 \cr
\check\sigma_{yx}(x,y) & \check\sigma_{yy}(x,y) & 0 \cr
         0 & 0 & 0
\end{array}
	\right)\quad \mbox{on}\ \Sigma.
\end{equation}
%%%%%%%%%%%%%%%%%%%%%%
For the time being, $\bjs[\bfe_\parallel|_\Sigma]$ and $\bfe_\parallel$ are viewed as $3$-component vectors.

Next, we derive integral equations for $\bfe_\parallel|_\Sigma$. Let $(\bfe, \bfb)$ be the electromagnetic field in $\mathbb{R}^3$. We use the vector potential, $\bfa$, where $\bfb=\nabla\times \bfa$ in the Lorentz gauge. Suppose $(\bfe^{\rm pr}, \bfb^{\rm pr})$ is the primary field, without the sheet ($\bjs=0$). By singling out the scattered field, $(\bfe^{\rm sc}, \bfb^{\rm sc})$, and its vector potential, $\bfa^{\rm sc}$, due to $\bjs$, we have~\cite{King-book}
%%%%%%%%%%%%%%%%%%%%%%
\begin{equation}\label{eq:A-sc-gen}
\bfa^{\rm sc}(\br)=\mu \Int\limits_\Sigma G(\br;\br')\,\bjs(\br')\,{\rm d}\br',
\end{equation}
%%%%%%%%%%%%%%%%%%%%%%
where $G$ is the Green function for the Helmholtz equation, viz., $(\Delta+k_0^2)G(\br;\br')=-\delta(\br-\br')$ in $\mathbb{R}^3$, where $k_0=\omega\sqrt{\varepsilon\mu}$, $\delta(\br)$ is the Dirac mass, and $\Delta$ is the 3D Laplacian. From the Amp\`ere-Maxwell law, the scattered electric field outside $\Sigma$ is obtained by
%%%%%%%%%%%%%%%%%%%%%%
\begin{equation*}
\bfe-\bfe^{\rm pr}=\bfe^{\rm sc}=\frac{\iu}{\omega\varepsilon\mu}\nabla\times\nabla\times\bfa^{\rm sc}.
\end{equation*}
%%%%%%%%%%%%%%%%%%%%%%
By projecting the electric field onto the $xy$-plane and letting $z$ approach 0, we find
%%%%%%%%%%%%%%%%%%%%%%
\begin{eqnarray}\label{eq:E-int_eq-gen}
\bfe_\parallel(\br)&=&\bfe_\parallel^{\rm pr}(\br)+\frac{\iu\omega\mu}{k_0^2} (\mathbf{I}-\ez\otimes \ez)\cdot\left\{\nabla\times\nabla \times \right.\nonumber \\
&&  \times \left. \Int\limits_\Sigma G(\br;\br')\,\bjs[\bfe_\parallel|_\Sigma](\br')\,{\rm d}\br'\right\},\quad \br=(x,y,0),
\end{eqnarray}
%%%%%%%%%%%%%%%%%%%%%%
where $\mathbf{I}$ is the $3\times 3$ unit matrix and $\otimes$ denotes a tensor product. This integral equation is applied to flat sheets, $\Sigma$, of arbitrary shapes, and is a particular case of the `electric field integral equation' formalism of  electromagnetics~\cite{Chew-book}. 

We now introduce the EPP with wave number $q$ in the $y$-direction ($q\in\mathbb{C}$) with the goal to determine $q$ so that~(\ref{eq:E-int_eq-gen}) has nontrivial solutions for zero incident field, if $\Sigma$ is a half plane (figure~\ref{fig:geom}). We take the following steps: (i) Assume $(\bfe^{\rm pr},\bfb^{\rm pr})=0$. (ii) Apply the ansatz $\boldsymbol{F}(x,y,z)=e^{\iu q y} \widetilde {\boldsymbol F}(x,z)$ in $\mathbb{R}^3$ with $\bjs[\bfe_\parallel|_\Sigma](x,y)=e^{iqy}\boldsymbol{\widetilde{\mathcal J}}[\widetilde{\bfe}_\parallel|_\Sigma](x)$ ($\boldsymbol F=\bfe, \bfb, \bfa$). Subsequently, we drop the tildes in this context for ease of notation.

After some  algebra, and writing $\bfe_\parallel(x)$ instead of $\bfe_\parallel(x,0)$, by~(\ref{eq:E-int_eq-gen}) we obtain
%%%%%%%%%%%%%%%%%%%%%%
\begin{eqnarray}\label{eq:int-eqn-sys-e}
\bfe_\parallel(x)&=&\frac{\iu\omega\mu}{k_0^2}
\left(\begin{array}{lr}
\displaystyle \frac{{\rm d}^2}{\dx^2}+k_0^2 &  \quad \displaystyle \iu q \frac{{\rm d}}{\dx}  \cr
  \displaystyle \iu q\frac{{\rm d}}{\dx}      &   k_0^2-q^2
  \end{array}
  \right) \nonumber\\
  && \mbox{} \times \int_0^\infty K(x-x')\, \bjs[\bfe_\parallel|_\Sigma](x')\,\dx',\ -\infty<x<\infty.
  \end{eqnarray}
  %%%%%%%%%%%%%%%%%%%%%%
 In this expression, we view $\bfe_\parallel$ and $\bjs(\bfe_\parallel|_\Sigma)$ as $2\times 1$ vectors, suppressing their (zero) $z$-component for simplicity, and the kernel is
 %%%%%%%%%%%%%%%%%%%%%%
 \begin{equation}\label{eq:kernel-hom}
 K(x)=\textstyle{\frac{\iu}{4}}H_0^{(1)}(\ke|x|),\quad \ke=\sqrt{k_0^2-q^2},\ \Im\,\ke>0,
 \end{equation}
 %%%%%%%%%%%%%%%%%%%%%%
 where $H_\nu^{(1)}(w)$ is the first-kind Hankel function of the $\nu$-th order~\cite{Bateman-II}. Note that this kernel incorporates the appropriate  radiation condition~\cite{King-book}.

We proceed to motivate the separation of two length scales. Suppose that $\sigma_{\#}$ is a scalar, positive function of $\omega$ having typical values of $|\bjs|/|\bfe_\parallel|$ on $\Sigma$. For example, $\sigma_{\#}=\sup_{\bfe_\parallel}\{|\bjs[\bfe_\parallel]|/|\bfe_\parallel|\}$ for all $\bfe_\parallel\neq 0$ on $\Sigma$. (We remark in passing that $\sigma_{\#}$ is the `operator norm' for $\bjs$). This $\sigma_{\#}$ has units of conductivity. We enforce the conditions
%%%%%%%%%%%%%%%%%%%%%%
 \begin{equation*}
 |q|\gg |k_0|\ \mbox{and}\ \left|\frac{\omega\mu\sigma_{\#}}{k_0}\right|\ll 1\qquad (\mbox{nonretarded\ regime}).
 \end{equation*}
 %%%%%%%%%%%%%%%%%%%%%%
These two conditions are interrelated, since $q$ should depend on $\sigma_{\#}$. Now define
%%%%%%%%%%%%%%%%%%%%%%
 \begin{equation*}
 l=	\frac{\omega\mu\sigma_{\#}}{k_0^2}
 \end{equation*}
 %%%%%%%%%%%%%%%%%%%%%%
which has units of length ($|k_0 l|\ll 1$). For example, if $\Sigma$ is isotropic and homogeneous with conductivity $\sigma$, $\Im\,\sigma>0$, then $\sigma_{\#}=|\sigma|$ and $l$ is of the order of the SPP wavelength~\cite{Bludov2013}. Since the EPP and SPP wave numbers may be comparable~\cite{Feietal2015}, take
%%%%%%%%%%%%%%%%%%%%%%
\begin{equation*}
|q l|= \mathcal O(1)\ \mbox{as}\ k_0 l \to 0.
\end{equation*}
%%%%%%%%%%%%%%%%%%%%%%

In view of the above, consider the formal two-scale expansion
%%%%%%%%%%%%%%%%%%%%%%
\begin{equation*}
\bfe_\parallel(x)= \mte^{(0)}(\brx,\bx)+(k_0 l)\mte^{(1)}(\brx,\bx)+\ldots,\quad \brx=\frac{x}{l},\ \bx=k_0 x,
\end{equation*}
%%%%%%%%%%%%%%%%%%%%%%
where $\brx$ is the fast scale, controlled by $l$, $\bx$ is the slow scale, set by the ambient-medium wavelength, and $k_0 l$ is the expansion parameter. The operator
${\rm d}/\dx$ is replaced by
%%%%%%%%%%%%%%%%%%%%%%
\begin{equation*}
\frac{{\rm d}}{\dx}\to	l^{-1}\frac{\partial}{\partial \brx}+k_0 \frac{\partial}{\partial \bx}.
\end{equation*}
%%%%%%%%%%%%%%%%%%%%%%
Let $\epsilon=k_0 l$ ($|\epsilon|\ll 1$). The substitution of the two-scale expansion into~(\ref{eq:int-eqn-sys-e}) yields
%%%%%%%%%%%%%%%%%%%%%%
\begin{eqnarray*}
\lefteqn{\mte^{(0)}(\brx,\bx)+\epsilon\,\mte^{(1)}(\brx,\bx)+\ldots +\epsilon^n \mte^{(n)}(\brx,\bx)+\ldots}\nonumber\\
&=&
\left(
\begin{array}{lr}
\displaystyle \frac{\partial^2}{\partial \brx^2}+2\epsilon\frac{\partial^2}{\partial\brx\partial\bx}+\epsilon^2\biggl(\frac{\partial^2}{\partial\bx^2}+1\biggr) & \displaystyle \iu (q l)\biggl(\frac{\partial}{\partial\brx}+\epsilon\frac{\partial}{\partial \bx}\biggr) \cr
\displaystyle \iu (q l)\biggl(\frac{\partial}{\partial\brx}+\epsilon\frac{\partial}{\partial\bx}\biggr) & -(q l)^2+\epsilon^2
\end{array}
\right) \nonumber\\
 &&  \times \int_0^\infty \mathcal K(\brx-\brx',\bx-\bx')
 \frac{\bjs[\mte^{(0)}]+\epsilon \bjs[\mte^{(1)}]+\ldots + \epsilon^n \bjs[\mte^{(n)}]+\ldots}{-\iu \sigma_{\#}}\,{\rm d}\brx'
 \end{eqnarray*}
 %%%%%%%%%%%%%%%%%%%%%%
where $\mathcal K(\brx,\bx)=(\iu/4)H_0^{(1)}(\sqrt{\bx^2-(q l)^2\brx^2})$ and $\bjs[\cdot]$ depends explicitly only on the fast variable. In the integral,
$\mathcal K(\brx-\brx',\epsilon(\brx-\brx'))$ must be expanded in powers of $\epsilon$ while $\brx-\brx'$ is treated as fixed.
By balancing out terms on both sides, we can in principle derive a cascade of equations for $\mte^{(n)}$ ($n=0,1,\ldots$). For $n=0$, we find
%%%%%%%%%%%%%%%%%%%%%%
\begin{eqnarray*}
	\mte^{(0)}(\brx)&=&
	\left(
\begin{array}{lr}
\displaystyle -\partial/\partial \brx & 0 \cr
\displaystyle \quad 0 & -\iu (q l)
\end{array}
\right) \nonumber\\
&& \mbox{} \times 	
\left(
\begin{array}{lr}
\displaystyle -\iu (\partial/\partial \brx) & \displaystyle q l \cr
\displaystyle -\iu (\partial/\partial\brx)  & q l
\end{array}
\right)\cdot \int_0^\infty \mathcal K(\brx-\brx',0)\frac{\bjs[\mte^{(0)}|_\Sigma](\brx')}{\sigma_{\#}}\,{\rm d}\brx',
\end{eqnarray*}
%%%%%%%%%%%%%%%%%%%%%%
where, abusing notation slightly, we write $\mte^{(0)}(\brx)$ instead of $\mte^{(0)}(\brx,\bx)$ in order to stress that this zeroth-order solution depends only on the fast variable. Notice that $(\partial/\partial \brx, \iu ql)$ is the surface gradient, given the ansatz for the $y$-dependence of the fields. Thus, to leading order in $k_0 l$, restoring the notation for the electric field, we write
%%%%%%%%%%%%%%%%%%%%%%
\begin{eqnarray*}
	\bfe_\parallel(x)&\sim &\frac{\iu\omega\mu}{k_0^2}
\left(\begin{array}{lr}
\displaystyle {\rm d}/\dx &  \quad 0 \cr
  0     &   \iu q    
  \end{array}
  \right) \cdot 
\left(\begin{array}{lr}
\displaystyle {\rm d}/\dx & \iu q  \cr
  {\rm d}/\dx       &   \iu q    
  \end{array}
  \right) \nonumber\\
  && \mbox{} \times \int_0^\infty \mathfrak K(x-x')\, \bjs[\bfe_\parallel|_\Sigma](x')\,\dx',\ \mathfrak K(x)=\textstyle{\frac{1}{2\pi}}K_0(q\sg(q)|x|),
  \end{eqnarray*}
  %%%%%%%%%%%%%%%%%%%%%%
  where $K_\nu(w)$ is the third-kind modified Bessel function of the $\nu$-th order~\cite{Bateman-II}, and $\sg(q)=1$ if $\Re\,q>0$ while $\sg(q)=-1$ if $\Re\,q<0$. Hence,
  %%%%%%%%%%%%%%%%%%%%%%
  \begin{equation*}
  \bfe_\parallel(x)\sim -({\rm d}/\dx, \iu q)\varphi(x), 	
  \end{equation*}
  %%%%%%%%%%%%%%%%%%%%%%
  where $\varphi(x)$ is a {\em bounded and continuous} scalar function that satisfies
  %%%%%%%%%%%%%%%%%%%%%%
    \begin{equation}\label{eq:int-phi-gen}
  \varphi(x)=-\frac{\iu \omega\mu}{k_0^2} \left(\frac{{\rm d}}{\dx}, \iu q\right)
  \!\cdot \!\!\int_0^{\infty} \mathfrak K(x-x')\,\bjs\biggl[\biggl(-\frac{{\rm d}\varphi}{\dx},-\iu q\varphi\biggr)\biggr](x')\,\dx'	
  \end{equation}
  %%%%%%%%%%%%%%%%%%%%%%
for all $x$. This expression may be simplified once $\bjs[\cdot]$ is specified. Equation~(\ref{eq:int-phi-gen}) amounts to the quasi-electrostatic approximation (cf.~\cite{VolkovMikhailov1988}). For a broad review on the quasi-electrostatic approximation, see, e.g.,~\cite{Larsson2007}. Our task is to determine $q$ so that~(\ref{eq:int-phi-gen}) has nontrivial admissible solutions, $\varphi(x)$, when $\bjs[\bfe_\parallel|_\Sigma]$ is described in terms of a local, spatially homogeneous surface conductivity tensor. 
%This scale separation also applies to strips $\Sigma$, which have a finite width along the $x$-axis and infinite straight edges, by use of a finite integration range in~(\ref{eq:int-phi-gen}).

%%%%%%%%%%%%%%%%%%%%%%%%%%%%%%%%%%%%%%%%%%%%%%%%%%%%%%%%%%%%%%%
%%%%%%%%%%%%%%%%%%%%%%%%%%%%%%%%%%%%%%%%%%%%%%%%%%%%%%%%%%%%%%%
\section{Homogeneous sheet: Wiener-Hopf method and the notion of the index}
\label{sec:solution}
%%%%%%%%%%%%%%%%%%%%%%%%%%%%%%%%%%%%%%%%%%%%%%%%%%%%%%%%%%%%%%%
%%%%%%%%%%%%%%%%%%%%%%%%%%%%%%%%%%%%%%%%%%%%%%%%%%%%%%%%%%%%%%%

In this section, we solve integral equation~(\ref{eq:int-phi-gen}) for the potential $\varphi$ on homogeneous and anisotropic sheets, with conductivity kernel  $\check{\bsig}(\br)=\bsig^\Sigma\delta(x) \delta(y) $ in~(\ref{eq:sigma-nonlocal}). In this vein, we introduce the related concept of the index, $\nu_K$. We also discuss a possible meaning of this $\nu_K$,   and describe the bulk SPPs that emerge as residue contributions to a Fourier integral for $\varphi$ for zero index $\nu_K$. \color{black} Hence, we set
%%%%%%%%%%%%%%%%%%%%%%
\begin{equation}\label{eq:sigma-matrix}
	\bjs[\bfe_\parallel]=\bsig^\Sigma \cdot \bfe_\parallel,\quad
	\bsig^\Sigma=\left(\begin{array}{lr}
\sigma_{xx} & \sigma_{xy} \cr
\sigma_{yx} & \sigma_{yy}
\end{array}
	\right),
\end{equation}
%%%%%%%%%%%%%%%%%%%%%%
where the matrix $\bsig^\Sigma$ is spatially constant.
For non-active (i.e., lossy or lossless) 2D materials, this $\bsig^\Sigma$ must have a positive semidefinite Hermitian part. In addition, this $\bsig^\Sigma$ is subject to the Onsager reciprocity relations~\cite{Onsager1931-I,Ziman-book}.  \color{black}

%%%%%%%%%%%%%%%%%%%%%%%%%%%%%%%%%%%%%%%%
\subsection{Wiener-Hopf method and zero index}
\label{subsec:W_H-method}
%%%%%%%%%%%%%%%%%%%%%%%%%%%%%%%%%%%%%%%%
After integration by parts, (\ref{eq:int-phi-gen}) becomes
%%%%%%%%%%%%%%%%%%%%%%
\begin{eqnarray}\label{eq:phi-int-hom}
\varphi(x)&=& \frac{\iu \omega\mu}{k_0^2} \biggl(\frac{{\rm d}}{\dx},\ \iu q\biggr)\cdot \bsig^\Sigma \cdot
\left(
\begin{array}{l}
{\rm d}/\dx \cr
\iu q
\end{array}
\right)
\int_0^\infty \mathfrak K(x-x')\varphi(x')\,\dx'\nonumber\\
&& \mbox{} -\frac{\iu\omega\mu}{k_0^2} \biggl(\frac{{\rm d}}{\dx},\ \iu q\biggr)\cdot \bsig^\Sigma \cdot
\left(
\begin{array}{l}
1 \cr
0
\end{array}
\right)
\mathfrak K(x) \varphi_0,\ -\infty< x<\infty,
\end{eqnarray}
%%%%%%%%%%%%%%%%%%%%%%
where $\varphi_0=\lim_{x\downarrow 0}\varphi(x)$ (cf.~equation~(16)  in~\cite{VolkovMikhailov1988}) and $\varphi(x)$ is continuous and integrable on the real axis~\footnote[1]{On the conducting sheet ($x>0$), the integrability assumed for $\varphi(x)$ may be viewed as a consequence of the requirements for finite total surface charge and bounded surface current density normal to the edge. Since $\varphi_0$ is finite, the boundedness of $\varphi(x)$ at the edge from outside $\Sigma$ is implied by continuity.}. We will describe $q$ so that~(\ref{eq:phi-int-hom}) admits such solutions with $\varphi_0\neq 0$.

In order to apply the Wiener-Hopf method~\cite{Krein1962}, we define the Fourier transforms
%%%%%%%%%%%%%%%%%%%%%%
\begin{equation*}
	\hatp(\xi)=\int_{-\infty}^\infty e^{-\iu \xi x}\varphi(x)\,\dx,\quad
	\hatp_{\pm}(\xi)=
	\int_{-\infty}^{\infty} e^{-\iu \xi x}\,\theta(\mp x)\varphi(x)\,\dx\ (\xi\in\mathbb{C})
\end{equation*}
%%%%%%%%%%%%%%%%%%%%%%
where $\theta(x)=1$ if $x>0$ and $\theta(x)=0$ if $x<0$. Thus, $\hatp(\xi)=\hatp_+(\xi)+\hatp_-(\xi)$ where $\hatp_\pm(\xi)$ is analytic in the upper ($+$) or lower ($-$) half plane, and $\hatp_\pm(\xi)\to 0$ as $\xi\to\infty$.

The application of the Fourier transform to~(\ref{eq:phi-int-hom}) yields
%%%%%%%%%%%%%%%%%%%%%%
\begin{equation}\label{eq:func-eq-xi}
\hatp_+(\xi)+\mathcal P(\xi)\,\hatp_-(\xi)=\frac{\omega\mu}{k_0^2}
(\sigma_{yx} q+\sigma_{xx}\xi)\hatK(\xi)\,\varphi_0,\ -\infty<\xi<\infty,	
\end{equation}
%%%%%%%%%%%%%%%%%%%%%%
where $\varphi_0=\lim_{x\downarrow 0}\varphi(x)=\lim_{x\uparrow 0}\varphi(x)$ by the continuity of $\varphi(x)$, and~\footnote{The function $\mathcal P(\xi)$ defined in~(\ref{eq:P-def}) should {\em not} be confused with a polynomial in $\xi$.}
%%%%%%%%%%%%%%%%%%%%%%
\begin{equation*}
 	\hatK(\xi)=\int_{-\infty}^\infty e^{-\iu \xi x} \mathfrak K(x)\,\dx=\textstyle{\frac{1}{2}}(\xi^2+q^2)^{-1/2},\quad \Re\sqrt{\xi^2+q^2}>0,
\end{equation*}
%%%%%%%%%%%%%%%%%%%%%%
\begin{equation}\label{eq:P-def}
	\mathcal P(\xi)=1+\frac{\iu\omega\mu}{k_0^2}\{\sigma_{xx}\xi^2+(\sigma_{xy}+\sigma_{yx})q\xi+\sigma_{yy}q^2\}\hatK(\xi).
\end{equation}
%%%%%%%%%%%%%%%%%%%%%%
We seek nontrivial $\hatp_{\pm}(\xi)$ (if $\varphi_0\neq 0$) that solve~(\ref{eq:func-eq-xi}). The condition $\Re\sqrt{\xi^2+q^2}>0$ is dictated by the assumption of waves that decay in the $z$-direction, and defines the first Riemann sheet. Note that, consequently, $\sqrt{\xi^2+q^2}$ is an even function of $\xi$.

To derive formulas for $\hatp_\pm(\xi)$, we need to separate the terms in~(\ref{eq:func-eq-xi}) into two types: terms that are analytic in the upper half plane and terms that are analytic in the lower half plane~\cite{Krein1962,Masujima-book}.
For this purpose, we first assume that
%%%%%%%%%%%%%%%%%%%%%%
\begin{equation*}
	\mathcal P(\xi)\neq 0\ \mbox{for\ all\ real}\ \xi.
\end{equation*}
%%%%%%%%%%%%%%%%%%%%%%
\medskip

\noindent {\bf Definition 1.} {\em Consider the `Krein index', $\nu_K$, associated with the function $\mathcal P(\xi)$ on the real axis. This $\nu_K$ is defined as} \cite{Krein1962}
%%%%%%%%%%%%%%%%%%%%%%
\begin{eqnarray}\label{eq:index-def}
\nu_K &=&\frac{1}{2\pi \iu}\lim_{M\to +\infty}
\int_{-M}^M \frac{\mathcal P'(\xi)}{\mathcal P(\xi)} \dxi\nonumber\\
&=&\frac{1}{2\pi\iu}\lim_{M\to +\infty}\oint_{\Gamma_0(M)} \frac{{\rm d}\mathcal P}{\mathcal P}=\frac{1}{2\pi}\arg\mathcal P(\xi)\bigl|_{\xi=-\infty}^{+\infty}.
\end{eqnarray}
%%%%%%%%%%%%%%%%%%%%%%
\medskip

In definition~1, the prime indicates differentiation with respect to $\xi$ and $\Gamma_0(M)$ is the {\em oriented} curve that forms the image in the $\mathcal P$-plane of the interval $(-M, M)$ of the real axis in the $\xi$-plane, under the mapping $\xi\mapsto \mathcal P(\xi)$, as $M\to +\infty$. For the $\mathcal P(\xi)$ at hand, $\Gamma_0$ is a {\em closed contour} and $\nu_K$ is an integer ($\nu_K=0,\pm 1, \pm 2$ as noted in section~\ref{subsec:ET}). This $\nu_K$ is the winding number of $\Gamma_0$ with respect to the origin in the $\mathcal P$-plane, expresses the change of $(2\pi\iu)^{-1}\ln\mathcal P(\xi)$ between the extremities of the real axis in the $\xi$-plane, and depends on $\bsig^\Sigma$, $\omega$ and $q$. An interpretation of the index $\nu_K$ is discussed in section~\ref{subsec:ET}. Some properties of $\nu_K$ are outlined in~\ref{app:index-eval}.

To find the EPP dispersion relation, the suitable scenario is to set (see section~\ref{sec:EPP_dispersion})
%%%%%%%%%%%%%%%%%%%%%%
\begin{equation*}
\nu_K=0.
\end{equation*}
%%%%%%%%%%%%%%%%%%%%%%
For example, consider the case with $\sigma_{xy}+\sigma_{yx}=0$, when $\mathcal P(\xi)$ is an even function, and then treat the nonzero sum of $\sigma_{xy}$ and $\sigma_{yx}$ as a perturbation. \color{black} 
In the present case with a vanishing $\nu_K$, we can {\em directly} factorize $\mathcal P(\xi)$ on the real line, and obtain an EPP dispersion relation (section~\ref{sec:EPP_dispersion}). We need to determine `split' functions, $Q_\pm(\xi)$, analytic and single valued in the upper ($+$) or lower ($-$) half plane, such that
%%%%%%%%%%%%%%%%%%%%%%
\begin{equation}\label{eq:P_Q+-}
	Q(\xi)=\ln\mathcal P(\xi)=Q_+(\xi) + Q_-(\xi),\quad -\infty<\xi<\infty.
\end{equation}
%%%%%%%%%%%%%%%%%%%%%%
Since $\ln\mathcal P(\xi)$ is analytic and single valued on $\mathbb{R}$ for $\nu_K=0$, we apply the Cauchy integral formula along the boundary of a rectangle with its two sides parallel to the real axis, and thus obtain~\cite{Krein1962,Masujima-book}
%%%%%%%%%%%%%%%%%%%%%%
\begin{equation}\label{eq:Q+-}
	Q_\pm (\xi)=\pm \frac{1}{2\pi \iu}\int_{-\infty}^\infty\frac{Q(\xi')}{\xi'-\xi}\,\dxi',  \quad Q(\xi)=\ln\mathcal P(\xi)\quad (\pm \Im\,\xi>0).
\end{equation}
%%%%%%%%%%%%%%%%%%%%%%
%For real $\xi$, $Q_\pm(\xi)$ can be obtained from the above formula by analytic continuation.

We proceed to describe $\varphi(x)$ for $\nu_K=0$. Equation~(\ref{eq:func-eq-xi}) becomes
%%%%%%%%%%%%%%%%%%%%%%
\begin{equation*}
e^{-Q_+(\xi)}\hatp_+(\xi)+e^{Q_-(\xi)}\hatp_-(\xi)=\frac{\omega\mu}{k_0^2}
(q\sigma_{yx}+\xi\sigma_{xx})\hatK(\xi) e^{-Q_+(\xi)}\varphi_0.
\end{equation*}
%%%%%%%%%%%%%%%%%%%%%%
The next step is to find suitable split functions, $\Lambda_{\pm}(\xi)$, such that the right-hand side is expanded as $-\iu[\Lambda_+(\xi)+\Lambda_-(\xi)]\varphi_0$ where the factor of $\iu$ is used for later algebraic convenience. This splitting can be carried out as follows. By~(\ref{eq:P-def})~and~(\ref{eq:P_Q+-}), we obtain
%%%%%%%%%%%%%%%%%%%%%%
\begin{equation*}
\frac{\iu\omega\mu}{k_0^2}(q\sigma_{yx}+\xi\sigma_{xx})\hatK(\xi) e^{-Q_+(\xi)}=\left(\frac{\mathcal C^+}{\xi-\xi^+}+\frac{\mathcal C^-}{\xi-\xi^-}\right) \left\{e^{Q_-(\xi)}-e^{-Q_+(\xi)}\right\},
\end{equation*}
%%%%%%%%%%%%%%%%%%%%%%
where $\xi^\pm$ are the zeros of the polynomial $\sigma_{xx}\xi^2+(\sigma_{xy}+\sigma_{yx})q\xi+\sigma_{yy}q^2$, viz.,
%%%%%%%%%%%%%%%%%%%%%%
\begin{equation}\label{eq:zeros-polyn}
\xi^\pm = -q \frac{(\sigma_{xy}+\sigma_{yx})\pm \sg(q)\,\mathfrak D}{2\sigma_{xx}},\
\mathfrak D=\sqrt{(\sigma_{xy}+\sigma_{yx})^2-4\sigma_{xx}\sigma_{yy}}
\end{equation}
%%%%%%%%%%%%%%%%%%%%%%
with $\pm \Re\,\xi^\pm >0$ and
%%%%%%%%%%%%%%%%%%%%%%
\begin{equation}\label{eq:C-coeffs}
\mathcal C^{\pm}=\frac{1}{2} \left\{1\pm \frac{\sigma_{xy}-\sigma_{yx}}{\mathfrak D}\,\sg(q)\right\}.
\end{equation}
%%%%%%%%%%%%%%%%%%%%%%
Hence, by inspection we can take
%%%%%%%%%%%%%%%%%%%%%%
\begin{eqnarray}\label{eq:Lambda-def}
\Lambda_\pm(\xi)&=&\mp \frac{\mathcal C^{\pm}}{\xi-\xi^\pm}\left\{e^{\mp Q_\pm(\xi)}-e^{\mp Q_\pm(\xi^\pm)}\right\} \pm \frac{\mathcal C^{\mp}}{\xi-\xi^\mp} \nonumber\\
&& \mbox{} \qquad \times \left\{ e^{\pm Q_\mp(\xi^\mp)}-e^{\mp Q_{\pm}(\xi)}\right\}.
\end{eqnarray}
%%%%%%%%%%%%%%%%%%%%%%

Thus, (\ref{eq:func-eq-xi}) leads to the statement
%%%%%%%%%%%%%%%%%%%%%%
\begin{equation*}
e^{-Q_+(\xi)}\hatp_+(\xi)+\iu \Lambda_+(\xi)\varphi_0=-e^{Q_-(\xi)}\hatp_-(\xi)-\iu \Lambda_-(\xi)\varphi_0
\end{equation*}
%%%%%%%%%%%%%%%%%%%%%%
for $-\infty<\xi<\infty$. By analytic continuation to the complex $\xi$-plane, the functions of the two sides of this equation together define an entire function, $\mathfrak E(\xi)$, which is a polynomial. By the asymptotic behavior of $Q_\pm(\xi)$ (\ref{app:Q-asympt}) and the property $\varphi_\pm(\xi)\to 0$ as $\xi\to\infty$,  we assert that $\mathfrak E(\xi)= 0$. By Fourier inversion, we write
%%%%%%%%%%%%%%%%%%%%%%
\begin{eqnarray}\label{eqs:phi-invFT}
\varphi(x)&=& \frac{\varphi_0}{2\pi\iu}\int_{-\infty}^{\infty}\Lambda_+(\xi)\,e^{Q_+(\xi)}\,e^{\iu \xi x}\,\dxi \quad \mbox{if}\ x<0,\nonumber\\
\varphi(x)&=&\frac{\varphi_0}{2\pi \iu}\int_{-\infty}^\infty \Lambda_-(\xi) e^{-Q_-(\xi)} \,e^{\iu \xi x}\,\dxi \quad \mbox{if}\ x>0.
\end{eqnarray}
%%%%%%%%%%%%%%%%%%%%%%
We show that, for {\em zero} $\nu_K$, (\ref{eqs:phi-invFT}) are compatible with the continuity of $\varphi(x)$ across the edge if $q$ satisfies a transcendental equation which is the EPP dispersion relation  (see section~\ref{sec:EPP_dispersion}). In contrast, if $\nu_K\neq 0$ such a dispersion relation is not meaningful. For this latter case, see remarks 1-3 in section~\ref{subsec:ET}, remarks~4 and~5 in section~\ref{sec:EPP_dispersion}, and the analysis in~\ref{app:nonzero-index}. 

%%%%%%%%%%%%%%%%%%%%%%%%%%%%%%%%%%%%%%%%%%%%%%%%%%%%%%%%%%%%
\subsection{On the meaning of the index $\nu_K$}
\label{subsec:ET}
%%%%%%%%%%%%%%%%%%%%%%%%%%%%%%%%%%%%%%%%%%%%%%%%%%%%%%%%%%%%

Next, we make an attempt to discuss an interpretation of the index $\nu_K$, which in our framework provides a quantitative criterion for the solvability of~(\ref{eq:phi-int-hom}) and, thus, for the existence of the EPP dispersion relation. An indication that such a criterion underlies the EPP can be traced in~\cite{VolkovMikhailov1988}, where the authors calculate the energy of edge magnetoplasmons in semi-infinite conducting sheets under a static magnetic field, using an antisymmetric tensor conductivity in the collisionless regime. In this work~\cite{VolkovMikhailov1988}, it is pointed out that a branch of the solution $\omega(q)$ from the EPP dispersion relation can become equal to the energy of the corresponding bulk SPP in the direction of the edge. When this occurs, the EPP energy $\omega$ versus the wave number $q$ enters a continuum region and, hence, the EPP dispersion relation is not meaningful~\cite{VolkovMikhailov1988}. We explore this transition systematically by use of the index $\nu_K$ and numerics in~\cite{MSLLM-preprint}.  

To further illustrate the concept of the index $\nu_K$ here, it is of interest to express this integer in terms of the number of zeros of the function $\mathcal P(\xi)$ defined by~(\ref{eq:P-def}). We can claim that, for a wide range of physical parameters, $\nu_K$ equals (see~\ref{app:index-eval})
%%%%%%%%%%%%%%%%%%%%%%%%%
\begin{equation}\label{eq:index-value}
\nu_K=\frac{N^+-N^-}{2}+\frac{N_*^+ -N^-_*}{2},
\end{equation}
%%%%%%%%%%%%%%%%%%%%%%%%%
where $N^\pm$ is the number of zeros of $\mathcal P(\xi)$ in the upper ($+$) or lower ($-$) half $\xi$-plane in the first Riemann sheet, and $N^\pm_*$ is the respective number with $\Re\sqrt{\xi^2+q^2}<0$, in the second Riemann sheet. Notice that $\nu_K=0,\pm 1, \pm 2$. Equation~(\ref{eq:index-value}) is corroborated by numerical examples (see remark~5 in section~\ref{sec:EPP_dispersion} for $\nu_K=0$, and discussion in~\ref{app:index-eval}). We herein propose formula~(\ref{eq:index-value}) as a conjecture motivated by our numerics.

In the remainder of this subsection, we discuss implications of~(\ref{eq:index-value}) by pointing out a connection of the nonretarded bulk SPPs, which propagate away form the edge, to the (zero or nonzero) index $\nu_K$. For fixed index $\nu_K$ and EPP wave number $q$, the zeros of $\mathcal P(\xi)$ in the top Riemann sheet amount to  bulk SPPs propagating in the $x$-direction for the respective {\em infinite} 2D material. This can be verified from~(\ref{eq:phi-int-hom}) by the replacement of the integration range by $(-\infty, \infty)$ and removal of the second line of this equation, since there is no edge. The application of the Fourier transform in $x$ to the resulting equation formally yields $\mathcal P(\xi)\hatp(\xi)=0$, which has nontrivial solutions $\hatp$ only at points $\xi_{\rm sp}^{(\ell)}$ in $\mathbb{C}$ with $\mathcal P(\xi_{\rm sp}^{(\ell)})=0$ and $\Re\sqrt{(\xi_{\rm sp}^{(\ell)})^2+q^2}>0$ ($\ell=1,\ldots, n$, $n\le 4$). These $\xi_{\rm sp}^{(\ell)}$ are $q$-dependent and amount to the wave numbers in the $x$-direction of bulk SPPs that have propagation constant equal to $q$ in the $y$-direction on the infinite 2D material. For an {\em anisotropic} material, the number of SPPs with $\Im\,\xi_{\rm sp}^{(\ell)}>0$, which propagate in the positive $x$-direction, may {\em not} in principle be equal to the number of SPPs with $\Im\,\xi_{\rm sp}^{(\ell)}<0$.  In this vein, the zeros of $\mathcal P(\xi)$ in the second Riemann sheet (where $\Re\sqrt{\xi^2+q^2}<0$) are wave numbers in the $x$-direction of surface wave modes for the infinite sheet that amount to exponentially growing solutions with $|z|$. \color{black}
\medskip

\noindent {\bf Remark 1.} Suppose that  no bulk wave modes are possible that are exponentially growing with $|z|$ \color{black} (thus, $N^+_*=0=N^-_*$). By~(\ref{eq:index-value}), we infer that, for a given EPP wave number $q$ along the edge, the index $\nu_K$ expresses half the difference between the numbers of modes for bulk SPPs that propagate in the positive and negative $x$-directions on the corresponding infinite flat sheet. Hence, if $\nu_K=0$ in this setting, when the EPP dispersion relation exists  (see section~\ref{sec:EPP_dispersion}), an equal number of bulk SPPs can propagate to the left and to the right of the edge on the respective infinite sheet.
\medskip

\noindent {\bf Remark 2.} More generally, for a given $q$ formula (\ref{eq:index-value}) implies that the index $\nu_K$ equals half the difference between the numbers of all, exponentially decaying and growing with $|z|$, types of bulk wave modes for the corresponding infinite flat sheet. \color{black}
\medskip

Among the (exponentially decaying with $|z|$) bulk SPPs associated with the zeros of $\mathcal P(\xi)$, only those with $\Im\,\xi_{\rm sp}^{(\ell)}>0$ manifest in our setting of a semi-infinite sheet, as we show in section~\ref{subsec:SPPs-zero-index} via contour integration for $x>0$.
By inspection of $\mathcal P(\xi)$, we see that the reflection of $q$ ($q\rightarrow -q$) results in the reflection of $\xi_{\rm sp}^{(\ell)}$ ($\xi_{\rm sp}^{(\ell)}\to -\xi_{\rm sp}^{(\ell)})$ for all $\ell=1,\ldots, n$. This reflection property can be extended to the corresponding bulk wave modes that grow exponentially with $|z|$. \color{black}
\medskip

\noindent {\bf Remark 3.} Hence, for a given EPP wave number, $q$, along the edge of the semi-infinite sheet, the index $\nu_K$ equals half the difference between the numbers of all (exponentially decaying and growing with $|z|$) types of bulk wave modes having wave number equal to $q$ in the $y$-direction and the respective modes that would ensue if the EPP reversed its direction of propagation. \color{black}

%%%%%%%%%%%%%%%%%%%%%%%%%%%%%%%%%%%%%%%%%%%%%%%%%%%%%%%%%%%%
\subsection{Bulk SPPs for zero index}
\label{subsec:SPPs-zero-index}
%%%%%%%%%%%%%%%%%%%%%%%%%%%%%%%%%%%%%%%%%%%%%%%%%%%%%%%%%%%%

For $\nu_K=0$ and a given $q$, we now describe the bulk SPPs propagating away from the edge via the requisite Fourier integral for $\varphi(x)$ ($x>0$). By~(\ref{eqs:phi-invFT}) and~(\ref{eq:P_Q+-}), we have
%%%%%%%%%%%%%%%%%%%%%%%%%%%%%%%%%%%%%%%%%%%%%%%%%%%%%%%%
\begin{eqnarray*}
\varphi(x)&=& \varphi_0\left\{\mathcal C^+ e^{\iu \xi^+ x}-\frac{1}{2\pi\iu} \int_{-\infty}^\infty \dxi\, e^{\iu \xi x}\left[\frac{\mathcal C^-}{\xi-\xi^-} e^{Q_-(\xi^-)} \right.\right.\nonumber\\
&& \mbox{} \left.\left. \qquad +\frac{\mathcal C^+}{\xi-\xi^+}e^{-Q_+(\xi^+)}\right]  \frac{e^{Q_+(\xi)}}{\mathcal P(\xi)} \right\},\quad x>0.
\end{eqnarray*}
%%%%%%%%%%%%%%%%%%%%%%%%%%%%%%%%%%%%%%%%%%%%%%%%%%%%%%%%
By closing the path in the upper half $\xi$-plane of the top Riemann sheet, we may have contributions from (i) residues of the integrand at the poles $\xi=\xi^+, \xi_{\rm sp}^{(l)}$ with $\Im\,\xi_{\rm sp}^{(l)}>0$, and (ii) the branch cut for $\xi=\iu q\sg(q)$. The residues give
%%%%%%%%%%%%%%%%%%%%%%%%%%%%%%%%%%%%%%%%%%%%%%%%%%%%%%%%
\begin{eqnarray*}
\varphi^{\rm res}(x)&=&\varphi_0 \mathcal C^+ (\bar \xi^+-\bar \xi^-) e^{-Q_+(k_{\rm sp}\bar\xi^+)}\nonumber\\
&&\qquad \times \sum_{\ell=1\atop \Im\,\xi_{\rm sp}^{(\ell)} >0}^n
\frac{e^{Q_+(k_{\rm sp}\bar\xi_\ell)} e^{\iu k_{\rm sp}\bar\xi_\ell x} \sqrt{\bar\xi_\ell^2+\bar q^2}}{\sqrt{\bar\xi_\ell^2+\bar q^2}\big[2\bar\xi_\ell-(\bar\xi^+ +\bar\xi^-)\big]-\bar\xi_\ell}
\end{eqnarray*}
%%%%%%%%%%%%%%%%%%%%%%%%%%%%%%%%%%%%%%%%%%%%%%%%%%%%%%%%
where $k_{\rm sp}=-2k_0^2/(\iu \omega\mu\sigma_{xx})$, the wave number of the transverse magnetic SPP~\cite{Bludov2013}, and $q=k_{\rm sp}\bar q$,  $\xi^{\pm}=k_{\rm sp}\bar\xi^{\pm}$ and $\xi_{\rm sp}^{(\ell)}=k_{\rm sp}\bar\xi_\ell$. In the nonretarded regime, we have $|k_{\rm sp}|\gg |k_0|$, $q$ satisfies~(\ref{eq:EPP-dr}) and each of $\bar q$, $\bar\xi^{\pm}$ and $\bar\xi_l$ is, in principle, an $\mathcal O(1)$ quantity.

In the setting with a tensor valued surface conductivity, it is possible to have zero and nonzero values of the index $\nu_K$ by manipulating the sum of the off-diagonal elements of $\bsig^\Sigma$, $\omega$ and $q$. A few physical predictions are explored elsewhere~\cite{MSLLM-preprint}.

\color{black}

%%%%%%%%%%%%%%%%%%%%%%%%%%%%%%%%%%%%%%%%%%%%%%%%%%%%%%%%%%%%
%%%%%%%%%%%%%%%%%%%%%%%%%%%%%%%%%%%%%%%%%%%%%%%%%%%%%%%%%%%%
\section{EPP dispersion relation on homogeneous sheet under zero index}
\label{sec:EPP_dispersion}
%%%%%%%%%%%%%%%%%%%%%%%%%%%%%%%%%%%%%%%%%%%%%%%%%%%%%%%%%%%%
%%%%%%%%%%%%%%%%%%%%%%%%%%%%%%%%%%%%%%%%%%%%%%%%%%%%%%%%%%%%
In this section, we derive the EPP dispersion relation from~(\ref{eqs:phi-invFT}), {\em if the index $\nu_K$ is zero}. We also summarize the theory pertaining to nonzero $\nu_K$ (see remark~4 and \ref{app:nonzero-index}). In addition, we provide examples of solutions of the derived EPP dispersion relation and the corresponding zero values of $\nu_K$ (see remark~5). \color{black}

First, we show by contour integration that the $\varphi(x)$ on the sheet given by~(\ref{eqs:phi-invFT}) is consistent with the definition of $\varphi_0$ without any further restriction of $q$. We compute
%%%%%%%%%%%%%%%%%%%%%%
\begin{eqnarray*}
\lefteqn{\lim_{x\downarrow 0}\varphi(x)=\lim_{\epsilon\downarrow 0}	\frac{\varphi_0}{2\pi\iu}\int_{-\infty}^\infty \Lambda_-(\xi) e^{-Q_-(\xi)} e^{\iu \xi\epsilon}\,\dxi  }\nonumber\\
 &=& \mathcal C^+\varphi_0 -\frac{\varphi_0}{2\pi\iu}\int_{-\infty}^\infty \left\{ \frac{\mathcal C^+}{\xi-\xi^+}e^{-Q_+(\xi^+)}
+\frac{\mathcal C^-}{\xi-\xi^-} e^{Q_-(\xi^-)}\right\}e^{-Q_-(\xi)}\,\dxi\nonumber\\
&=& (\mathcal C^+ + \mathcal C^-)\varphi_0 =\varphi_0,
\end{eqnarray*}
%%%%%%%%%%%%%%%%%%%%%%
where we use~(\ref{eq:Lambda-def}) for $\Lambda_\pm(\xi)$, a result for $Q_-(\xi)$ from~\ref{app:Q-asympt} and~(\ref{eq:C-coeffs}) for $\mathcal C^\pm$.

Next, we resort to the limit value of $\varphi(x)$ as $x$ approaches the edge from {\em outside} the sheet. Therefore, we calculate
%%%%%%%%%%%%%%%%%%%%%%
\begin{eqnarray*}
	\lim_{x\uparrow 0}\varphi(x)&=&I=\lim_{\epsilon\downarrow 0}	\frac{\varphi_0}{2\pi\iu}\int_{-\infty}^\infty \Lambda_+(\xi) e^{Q_+(\xi)} e^{-\iu \xi\epsilon}\,\dxi  \nonumber\\
	&=& \mathcal C^-\varphi_0 + \lim_{\epsilon\downarrow 0}\frac{\varphi_0}{2\pi\iu}\int_{-\infty}^\infty \dxi\ \left\{\frac{\mathcal C^+}{\xi-\xi^+} e^{-Q_+(\xi^+)}+\frac{\mathcal C^-}{\xi-\xi^-}e^{Q_-(\xi^-)}\right\} \nonumber\\
	&& \mbox{} \hphantom{\mathcal C^- + \lim_{\epsilon\downarrow 0}\frac{1}{2\pi\iu}\int_{-\infty}^\infty \dxi} \times e^{-Q_-(\xi)} \mathcal P(\xi) e^{-\iu\xi\epsilon},
\end{eqnarray*}
%%%%%%%%%%%%%%%%%%%%%%
using~(\ref{eq:Lambda-def}) and $e^{Q_+(\xi)}=e^{-Q_-(\xi)}\mathcal P(\xi)$. By~(\ref{eq:P-def}) and~(\ref{eq:zeros-polyn}) for $\mathcal P(\xi)$ and $\xi^\pm$, we have
%%%%%%%%%%%%%%%%%%%%%%
\begin{eqnarray*}
I&=&\frac{\iu \omega\mu\sigma_{xx}}{2k_0^2} \left[\mathcal C^+ e^{-Q_+(\xi^+)}+\mathcal C^- e^{Q_-(\xi^-)}\right]\lim_{\epsilon\downarrow 0}\frac{\varphi_0}{2\pi\iu}\int_{-\infty}^\infty \xi \frac{e^{-Q_-(\xi)}}{\sqrt{\xi^2+q^2}} e^{-\iu \xi\epsilon}\,\dxi \nonumber\\
 && \mbox{} -\frac{\iu\omega\mu\sigma_{xx}}{2k_0^2}\left[\mathcal C^+ \xi^- e^{-Q_+(\xi^+)}+\mathcal C^- \xi^+ e^{Q_-(\xi^-)}\right]\frac{\varphi_0}{2\pi\iu}\int_{-\infty}^\infty \frac{e^{-Q_-(\xi)}}{\sqrt{\xi^2+q^2}}\,\dxi.
\end{eqnarray*}
%%%%%%%%%%%%%%%%%%%%%%

 The limit of the integral in the first line of the last equation is infinity, as we verify by wrapping the path around the branch cut that emanates from $\xi=-\iu q\sg(q)$. In contrast, the integral  of the second line is convergent.
To eliminate the above divergence, we require that the corresponding prefactor vanishes, which in turn implies
%%%%%%%%%%%%%%%%%%%%%%
\begin{equation}\label{eq:EPP-dr}
Q_-(\xi^-)+Q_+(\xi^+)=\ln\Biggl(-\frac{\mathcal C^+}{\mathcal C^-}\Biggr)\qquad (\pm \Re \xi^\pm >0).
\end{equation}
%%%%%%%%%%%%%%%%%%%%%%
This relation is a highlight of our analysis (see comments below). Recall that $\xi^\pm$ and $\mathcal C^{\pm}$ are defined by~(\ref{eq:zeros-polyn}) and~(\ref{eq:C-coeffs}). If $\sigma_{xy}=-\sigma_{yx}$ and $\sigma_{xx}=\sigma_{yy}$ then (\ref{eq:EPP-dr}) becomes
%%%%%%%%%%%%%%%%%%%%%%
\begin{equation*}
\frac{\iu \sigma_{yx}}{\sigma_{xx}}\sg(q) \tanh\Biggl[\frac{1}{\pi}\int_0^\infty \frac{ \ln\left(1+\frac{\iu\omega\mu\sigma_{xx}}{2k_0^2}\sg(q)q\sqrt{1+\zeta^2}\right)}{1+\zeta^2}\,\dzet\Biggr]+1=0, 	\end{equation*}
%%%%%%%%%%%%%%%%%%%%%%
in agreement with equation~(12) in~\cite{VolkovMikhailov1985} (for vanishing slab thickness), and equation (28) in~\cite{VolkovMikhailov1988} for a single conducting sheet.

We can now verify that $\varphi(x)$ tends to $\varphi_0$ as $x$ approaches the edge from outside the sheet. Evaluating the convergent integral in the last expression for $I$, we have
%%%%%%%%%%%%%%%%%%%%%%%
\begin{eqnarray*}
	\lefteqn{\frac{\iu\omega\mu\sigma_{xx}}{2k_0^2}\frac{1}{2\pi\iu}\int_{-\infty}^\infty \frac{e^{-Q_-(\xi)}}{\sqrt{\xi^2+q^2}}\,\dxi=\frac{1}{2\pi\iu}\int_{-\infty}^\infty \frac{e^{Q_+(\xi)}-e^{-Q_-(\xi)}}{(\xi-\xi^+) (\xi-\xi^-)}\,\dxi}\\
	&& \mbox{ }=\frac{e^{Q_+(\xi^+)}-e^{-Q_-(\xi^-)}}{\xi^+-\xi^-}
\end{eqnarray*}
%%%%%%%%%%%%%%%%%%%%%%%
which, after some algebra, indeed furnishes that $\varphi(x)\to \varphi_0$ as $x\uparrow 0$ by virtue of~(\ref{eq:EPP-dr}).

The following comments on~(\ref{eq:EPP-dr}) are in order. (i) Evidently, its left-hand side is invariant under the reflection of $q$ ($q\to -q$) while the right-hand side is {\em not} unless $\sigma_{xy}=\sigma_{yx}$, as expected by symmetry considerations~\cite{VolkovMikhailov1988}. (ii) Because $Q_\pm(\xi)$ is analytic and single valued in the upper ($+$) or lower ($-$) half plane for $\nu_K=0$, only one branch of the logarithms involved in the integrals and on the right-hand of~(\ref{eq:EPP-dr}) is relevant~\cite{VolkovMikhailov1988}. By considering the long-wavelength asymptotics (section~\ref{sec:low_freq}), we choose a branch of $w=\ln \Xi$ by taking $-\pi<\Im w\le \pi$ for~(\ref{eq:EPP-dr}), for any requisite complex variable $\Xi$ ($\Xi\in\mathbb{C}$). Accordingly, we have $\ln(-\mathcal C^+/\mathcal C^-)= 0$ if $\mathcal C^+/\mathcal C^-= -1$ on the right-hand side of~(\ref{eq:EPP-dr}) (see section~\ref{sec:low_freq}). \color{black} (iii) The validity of~(\ref{eq:EPP-dr}) is limited due to the underlying approximations
(section~\ref{sec:formulation-asympt}). 
By our formalism, we expect that~(\ref{eq:EPP-dr}) holds if $\omega\mu\sigma_\#/|k_0|\ll 1$ where, e.g., $\sigma_\#=\sqrt{|\sigma_{xx}|^2+|\sigma_{yy}|^2+|\sigma_{xy}|^2+|\sigma_{yx}|^2}$.
\medskip

\noindent {\bf Remark 4.} We summarize the results for {\em nonzero index}, $\nu_K\neq 0$ (see the analysis of~\ref{app:nonzero-index}). If $\nu_K<0$, there is no $q$ for which integral equation~(\ref{eq:phi-int-hom}) admits nontrivial integrable and continuous solutions. If $\nu_K >0$ there is a region for $(q, \omega)$, with a continuum of complex values of $q$ for every $\omega$, such that~(\ref{eq:phi-int-hom}) has nontrivial solutions. A more detailed study of these cases is the subject of a separate paper~\cite{MSLLM-preprint}.

\medskip

\noindent {\bf Remark 5.} We now give examples of solutions, $q$, to relation~(\ref{eq:EPP-dr}), and respective zero index $\nu_K$. Our starting point is a diagonal $\bsig^\Sigma$ of the Drude form with possibly unequal diagonal elements~\cite{Nemilentsau2016}. We generate a nonzero sum $\sigma_{xy}+\sigma_{yx}$ via rotation in the $xy$-plane~\footnote{This matrix transformation can be viewed as the result of applying rotation of the unit cell of the bulk 2D material and then cutting the material to create an edge along a fixed axis.}. For $\nu_K=0$, we self-consistently obtain $q$ by numerically solving~(\ref{eq:EPP-dr}) with $\sigma_{xy}=\sigma_{yx}$ which implies that if $q$ is a solution so is $-q$ (see cases (i)--(iii) of this remark below).  In nondimensional form, the function $\mathcal P(\xi)=\mathcal P(\xi;q)$ reads
%%%%%%%%%%%%%%%%%%%%%%%%%%%%%%%%%%%%%%%%%%%%%%%%%%%%%%%
\begin{equation*}
	\mathcal P(\xi)=1+\frac{\iu}{2} \frac{\bar\sigma_{xx}\bar\xi^2+(\bar\sigma_{xy}+\bar\sigma_{yx})\bar q\,\bar\xi+\bar\sigma_{yy}\bar q^2}{(\bar q^2+\bar\xi^2)^{1/2}},\quad \bar\zeta=\frac{\zeta}{k_0}, \ \bar\sigma_{\ell\ell'}=\frac{\sigma_{\ell\ell'}}{\sqrt{\varepsilon/\mu}}
\end{equation*}
%%%%%%%%%%%%%%%%%%%%%%%%%%%%%%%%%%%%%%%%%%%%%%%%%%%%%%%
for $\zeta=\xi, q$ ($\ell, \ell'=x,y$). Here, $\Re(\bar\xi^2+\bar q^2)^{1/2}>0$, the ambient medium is considered to be lossless ($k_0>0$), and the symbols $\bar q$ and $\bar\xi$ should not be confused with similar symbols used in the end of section~\ref{subsec:ET}. If $\boldsymbol U(\phi)$ is the matrix representing the rotation by angle $\phi$ with respect to the positive $x$-axis in the $xy$-plane, we consider the matrix
%%%%%%%%%%%%%%%%%%%%%%%%%%%%%%%%%%%%%%%%%%%%%%%%%%%%%%%
\begin{equation*}
	\bar{\boldsymbol{\sigma}}^\Sigma(\phi)=\boldsymbol U(\phi) \bar{\boldsymbol{\sigma}}^\Sigma(0)\boldsymbol U(-\phi),\quad 0\le \phi<\pi,
\end{equation*}
%%%%%%%%%%%%%%%%%%%%%%%%%%%%%%%%%%%%%%%%%%%%%%%%%%%%%%%
for different choices of $\bar{\boldsymbol{\sigma}}^\Sigma(0)$ and angle $\phi$. Note that $\bar{\boldsymbol{\sigma}}^\Sigma(\phi)$ has elements $\bar\sigma_{\ell\ell'}(\phi)$.

\noindent {\em (i) Reference case: $\bar{\boldsymbol{\sigma}}^\Sigma(0)={\rm diag}(0.2\,\iu, 0.2\,\iu)$, $\phi=0$.} By solving~(\ref{eq:EPP-dr}) for $\Re\,q>0$ and finding $\nu_K$ as a winding number in the $\mathcal P$-plane by~(\ref{eq:index-def}), we compute $\bar q\simeq 12.172\simeq 1.217 k_{\rm sp}/k_0$, in agreement with~\cite{VolkovMikhailov1988}, and obtain $\nu_K=0$ as expected by~(\ref{eq:index-value}).

\noindent {\em (ii) $\bar{\boldsymbol{\sigma}}^\Sigma(0)={\rm diag}(0.001+0.1\iu, 0.002+0.2\iu)$, $\phi=0$.} The resulting $\mathcal P(\xi;q)$ is even in $\xi$ and has 4 zeros (symmetric with respect to the origin) in the top Riemann sheet. By~(\ref{eq:EPP-dr}) with $\Re\,q>0$ we find $\bar q\simeq 13.928+0.140 \iu$, and verify that $\nu_K=0$ directly by~(\ref{eq:index-def}), in agreement with~(\ref{eq:index-value}).

\noindent {\em (iii) $\bar{\boldsymbol{\sigma}}^\Sigma(0)={\rm diag}(0.001+0.1\iu, 0.002+0.2\iu)$, $\phi=0.4\pi$.} The ensuing $\mathcal P(\xi;q)$ has 2 zeros in the top Riemann sheet with $N^+=1=N^-$, while $N_*^+=1=N_*^-$. By~(\ref{eq:EPP-dr}) for $\Re\,q>0$ we obtain $\bar q\simeq 21.657+0.217 \iu$, and obtain $\nu_K=0$ by~(\ref{eq:index-def}), as found by~(\ref{eq:index-value}).
\medskip
\color{black}

%\noindent {\em (iv) $\bar{\boldsymbol{\sigma}}^\Sigma(0)={\rm diag}(0.001+0.1 \iu, 0.002+0.2\iu)$, $\phi=0.4\pi$ and $\bar q=0.850 \bar q^{(iii)}$.} In this case, we do not use the EPP dispersion relation. The resulting $\mathcal P(\xi;q)$ has 2 zeros in the lower half plane of the first Riemann sheet, with $N^+=0$ and $N^-=2$, while $N_*^+=N_*^-=1$. By~(\ref{eq:index-def}) we find $\nu_K=-1$, in agreement with~(\ref{eq:index-value}). By reflection of $q$, setting $q=-0.850 q^{(iii)}$, we verify that $\nu_K$ changes sign thus becoming $\nu_K=1$. 

%In the above, $N_*^+=N_*^-$. For a nonzero $N_*^+ - N_*^-$, consider the following steps.

%\noindent {\em (v) $\bar{\boldsymbol{\sigma}}^\Sigma(0)={\rm diag}(0.001+0.1 \iu, 0.002+0.2\iu)$ and $\phi=0.166\pi$.} First, we solve~(\ref{eq:EPP-dr}) and obtain $\bar q\simeq 16.438 + 0.164\iu=\bar q^{(v)}$, while we find $\nu_K=0$ by definition~(\ref{eq:index-def}). Then, we scale this $\bar q$ by taking $\bar q= 0.75 q^{(v)}$, thus abandoning relation~(\ref{eq:EPP-dr}). The ensuing $\mathcal P(\xi;q)$ has 3 zeros in the first Riemann sheet with $N^+=0$ and $N^-=3$, and only 1 zero in the second Riemann sheet with $N_*^+=1$ (and $N_*^-=0$). By~(\ref{eq:index-def}) we directly find $\nu_K=-1$, in agreement with formula~(\ref{eq:index-value}).

We conclude this section by noting that one may explore the values of $\nu_K$ in the parameter space of arbitrary $\bsig^\Sigma$, by invoking nonzero $\sigma_{xy}-\sigma_{yx}$ and $\sigma_{xy}+\sigma_{yx}$. For example, one can start with having $\sigma_{xx}\neq \sigma_{yy}$ and a static magnetic field~\cite{Fetter1985}, apply rotations and vary the applied magnetic field. Numerically, a transition from zero to nonzero $\nu_K$ occurs when it is no longer possible to determine a solution $q(\omega)$ to~(\ref{eq:EPP-dr}).

%%%%%%%%%%%%%%%%%%%%%%%%%%%%%%%%%%%%%%%%%%%%%%%%%%%%%%%%%%%%
%%%%%%%%%%%%%%%%%%%%%%%%%%%%%%%%%%%%%%%%%%%%%%%%%%%%%%%%%%%%
\section{Long-wavelength asymptotic expansion under zero index}
\label{sec:low_freq}
%%%%%%%%%%%%%%%%%%%%%%%%%%%%%%%%%%%%%%%%%%%%%%%%%%%%%%%%%%%%
%%%%%%%%%%%%%%%%%%%%%%%%%%%%%%%%%%%%%%%%%%%%%%%%%%%%%%%%%%%%
In this section, we derive an asymptotic expansion for $q$ from dispersion relation~(\ref{eq:EPP-dr}) in the regime described by $|\omega\mu\sigma_\#/k_0|\ll 1$  along with
(see sections~\ref{sec:formulation-asympt} and~\ref{sec:EPP_dispersion})
%%%%%%%%%%%%%%%%%%%%%%
\begin{equation}\label{eq:q-low_freq-asympt}
|k_0|\ll |q|< \mathcal O\biggl(\frac{2k_0^2}{\omega\mu\sigma_{\#}}\biggr),\quad 0\neq \frac{\sigma_{xy}-\sigma_{yx}}{\sigma_{xx}}=\mathcal O(1).
\end{equation}
%%%%%%%%%%%%%%%%%%%%%%
The above inequalities involving $|q|$ imply that the EPP wave number, $q$, is large in magnitude compared to the ambient medium propagation constant but much smaller
than the typical SPP wave numbers for $q=0$ on the sheet. Hence,  $|q l|$ is small compared to unity but much larger than $|k_0 l|$ ($|k_0 l|\ll 1$) where $l=\omega\mu\sigma_{\#}/k_0^2$. We also discuss the case with $\sigma_{xy}=\sigma_{yx}$, if the second condition in~(\ref{eq:q-low_freq-asympt}) does not hold. \color{black} 

First, let us examine the regime described by (\ref{eq:q-low_freq-asympt}). \color{black} By manipulating the integrals for $Q_+(\xi^+)$ and $Q_-(\xi^-)$ with $\xi^\pm=q\alpha^\pm$, we obtain
%%%%%%%%%%%%%%%%%%%%%%
\begin{eqnarray}\label{eq:Q_low_freq}
	Q_\pm(\xi^\pm)&=& \pm \frac{1}{2\pi\iu}\int_0^\infty
	 \frac{\dzet}{\zeta^2-(\alpha^\pm)^2} \biggl\{\zeta\bigl[\ln(1+\bq A(\zeta))-\ln(1+\bq A(-\zeta))\bigr]   \nonumber\\
	&& \mbox{} +\alpha^\pm  \bigl[\ln(1+\bq A(\zeta))+\ln(1+\bq A(-\zeta))\bigr]\biggr\}\sg(q),
\end{eqnarray}
%%%%%%%%%%%%%%%%%%%%%%
where
%%%%%%%%%%%%%%%%%%%%%%
\begin{equation*}
\bq=\frac{\iu \omega\mu\sigma_{xx}}{2k_0} \frac{q\sg(q)}{k_0},\
\alpha^\pm= -\frac{(\sigma_{xy}+\sigma_{yx})\pm \sg(q)\sqrt{(\sigma_{xy}+\sigma_{yx})^2-4\sigma_{xx}\sigma_{yy}}}{2\sigma_{xx}},	
\end{equation*}
%%%%%%%%%%%%%%%%%%%%%%
\begin{equation*}
	A(\zeta)=(\zeta-\alpha^+)(\zeta-\alpha^-) (1+\zeta^2)^{-1/2},\ \Re\left\{q{\rm sg}(q)\sqrt{1+\zeta^2}\right\}>0.
\end{equation*}
%%%%%%%%%%%%%%%%%%%%%%
We consider $|\alpha^\pm|=\mathcal O(1)$ and $|\bq|\ll 1$.
In the limit $\bq\to 0$, each of $Q_+(\xi^+)$ and $Q_-(\xi^-)$ approaches zero. This occurs since the branch of the logarithm $w=\ln\Xi$ is defined so that $-\pi<\Im w\le \pi$ which in turn implies $\ln(1)=0$, where $\Xi$ may be a function of $\zeta$ as seen by the above integrals for $Q_\pm(\xi^\pm)$. \color{black} 
Thus, for consistency of our assumptions for $|q|$ in this regime with dispersion relation~(\ref{eq:EPP-dr}), the ratio $-\mathcal C^+/\mathcal C^-$ must be close to unity. In view of~(\ref{eq:C-coeffs}), this implies that $(\sigma_{xy}-\sigma_{yx})/\mathfrak D$ should be large in magnitude. This behavior can characterize, for example, the conductivity $\bsig^\Sigma$ that comes from the hydrodynamic model in~\cite{Fetter1985} at low enough $\omega$ (see below).

We proceed to give the main result. By applying to $Q_\pm(\xi^\pm)$ the Mellin transform with respect to
$\bq$~\cite{SasielaShelton1993,Fikioris-asympt-book}, we obtain the low-order expansion (see~\ref{app:MT})
%%%%%%%%%%%%%%%%%%%%%%
\begin{equation}\label{eq:Q-exp-loworder}
	Q_+(\xi^+)+Q_-(\xi^-)\sim \frac{1}{\iu \pi} {\bq}\, (\alpha^+ -\alpha^-)\Biggl[
	\ln\biggl(\frac{2}{\bq}\biggr)+1 \Biggr]\ \mbox{as}\ \bq\to 0.
\end{equation}
%%%%%%%%%%%%%%%%%%%%%%
Accordingly, dispersion relation~(\ref{eq:EPP-dr}) becomes
%%%%%%%%%%%%%%%%%%%%%%
\begin{equation}\label{eq:q-longw}
	-\frac{1}{2\pi}\frac{\omega\mu (\sigma_{xy}-\sigma_{yx})}{2k_0}\frac{q}{k_0}\Biggl[
	\ln\biggl(\frac{2}{\bq}\biggr)+1 \Biggr]\simeq 1,
\end{equation}
%%%%%%%%%%%%%%%%%%%%%%
in agreement with equation~(41) in~\cite{VolkovMikhailov1988} for the special case with $\sigma_{xy}=-\sigma_{yx}$ and $\sigma_{xx}=\sigma_{yy}$. Our methodology can produce higher-order terms in~(\ref{eq:Q-exp-loworder}), which we omit here, but the procedure becomes increasingly cumbersome with the expansion order.

We further comment on~(\ref{eq:q-longw}). First, only $\sigma_{xx}$ and the ``antisymmetric part'' $\sigma_{xy}-\sigma_{yx}$ of $\bsig^\Sigma$ contribute to this long-wavelength expansion. Second, in order to extract $q(\omega)$ or $\omega(q)$ from~(\ref{eq:q-longw}), we would need to specify how the relevant matrix elements of $\bsig^\Sigma$ depend on $\omega$. For example, suppose that $\bsig^\Sigma$ emerges from a model based on linearized hydrodynamic equations for the electron flow in the presence of a static transverse magnetic field, $B_0 \ez$~\cite{Fetter1985,VolkovMikhailov1988,CohenGoldstein2018}. By neglecting the (nonlocal) effects of compressibility and viscosity, we apply a linear response theory and obtain
%%%%%%%%%%%%%%%%%%%%%%
\begin{equation}\label{eq:sigma-low_freq}
	\bsig^\Sigma(\omega)\simeq -\frac{\iu e^2 n_0}{\omega_c^2}
	 \left(
	\begin{array}{lr}
	\omega & -\iu \omega_c \\
	 \iu \omega_c & \omega
	\end{array}
	\right),\quad \tau^{-1}\ll \omega\ll |\omega_c|=\frac{e |B_0|}{m_e}.
\end{equation}
%%%%%%%%%%%%%%%%%%%%%%
Here, $e$ is the electron charge, $n_0$ is the background (unperturbed) electron density, $m_e$ is the electron mass, $\tau$ is a phenomenological relaxation time due to collisions of electrons with impurities or other particles, and $\omega_c=e B_0/m_e$ is the signed cyclotron `frequency'~\cite{Fetter1985}. In this formulation, we neglect the effect that the edge as a boundary of the 2D electron system has on the effective $\bsig^\Sigma$. Note that we should require that
%%%%%%%%%%%%%%%%%%%%%%
\begin{equation*}
	\biggl|\frac{\sigma_{xy}-\sigma_{yx}}{\mathfrak D}\biggr|\simeq \biggl|\frac{\omega_c}{\omega}\biggr|\gg 1,
\end{equation*}
%%%%%%%%%%%%%%%%%%%%%%
for compatibility with the regime of small $|\bq|$ that we study. Consequently, we find~\cite{VolkovMikhailov1988}
%%%%%%%%%%%%%%%%%%%%%%
\begin{equation*}
	\omega(q) \sim \frac{1}{2\pi}\frac{e^2n_0}{\varepsilon \omega_c}q\biggl[ \ln\left(\frac{4\varepsilon \omega_c^2}{e^2n_0 |q|}\right)+1\biggr]\quad \mbox{if}\ \tau^{-1}\ll \omega\ll |\omega_c|.
\end{equation*}
%%%%%%%%%%%%%%%%%%%%%%

Next, we discuss the distinctly different case with $\sigma_{xy}=\sigma_{yx}$. The term $\ln(-\mathcal C^+/\mathcal C^-)$ on the right-hand side of relation~(\ref{eq:EPP-dr}) then becomes equal to $\iu\pi$, since $\mathcal C^+=\mathcal C^-=1/2$. Hence, by inspection of the two sides of~(\ref{eq:EPP-dr}), we realize that the behavior of $q$ may be modified. If $|\alpha^{\pm}|$ are fixed then $|\breve q|$ cannot be small. By use  of~(\ref{eq:Q_low_freq}) in~(\ref{eq:EPP-dr}), we end up with a transcendental equation of the form 
%%%%%%%%%%%%%%%%%%%%%%
\begin{equation*}
\mathcal F(\breve q; \bar\sigma_1,\bar\sigma_2)=0\quad \mbox{where}\quad \mathcal F=Q_+(\xi^+)+Q_-(\xi^-)-\iu\pi, 
\end{equation*}
%%%%%%%%%%%%%%%%%%%%%%
with $\bar\sigma_1=(\sigma_{xy}+\sigma_{yx})/\sigma_{xx}$ and $\bar\sigma_2=\sigma_{yy}/\sigma_{xx}$.
The above equation may not be further simplified.
Any admissible solution $\breve q$ should be an $\mathcal O(1)$ quantity which depends on the parameters $\bar\sigma_1$ and $\bar\sigma_2$ entering $\alpha^\pm$. If $\breve q=\vartheta(\bar\sigma_1,\bar\sigma_2)$ denotes a solution, we have
%%%%%%%%%%%%%%%%%%%%%%
\begin{equation*}
q= \pm  \vartheta(\bar\sigma_1,\bar\sigma_2)\,\frac{\iu 2k_0^2}{\omega\mu\sigma_{xx}}~,\quad \vartheta=\mathcal O(1).
\end{equation*}
%%%%%%%%%%%%%%%%%%%%%%
This formula suggests that $q(\omega)=\mathcal O(\omega^2)$, if $\sigma_{xx}=\mathcal O(1/\omega)$ and $\bar\sigma_{1,2}=\mathcal O(1)$ as expected by the Drude model in the absence of a static magnetic field  for low frequency in the collisionless regime~\cite{Jablan2013,Lowetal2017,Nemilentsau2016}.
A numerical study of this case lies beyond our scope.

\color{black}

%In addition, bear in mind the conditions for the quasi-electrostatic approximation.
%Thus, bounds for $\omega$ or $|q(\omega)|$ may be found.

%A richer model for $\bsig^\Sigma$, with nonzero $\sigma_{xy}+\sigma_{yx}$, results from transforming a $\bsig^\Sigma$ for %which $\sigma_{xy}+\sigma_{yx}=0$ through arbitrary rotations  in the $xy$-plane~\cite{MSLLM-preprint} (see %section~\ref{subsec:EPP_dispersion_num}). Suppose that $\sigma_{xx}\neq \sigma_{yy}$ and $\sigma_{xy}=-\sigma_{yx}$ in the original $%\bsig^\Sigma$. According to our analysis, the action of rotation on this matrix is felt only to orders $\breve q^2$ and higher in the %long-wavelength expansion of~(\ref{eq:EPP-dr}).  The incorporation of viscosity or compressibility effects into $\bsig^\Sigma$ results %in nonlocal conductivity models, which will be the subject of future work. [{\bf DM: Tobias, Tony, any comments?}] \color{black}

%%%%%%%%%%%%%%%%%%%%%%%%%%%%%%%%%%%%%%%%%%%%%%%%%%%%%%%%%%%%
%%%%%%%%%%%%%%%%%%%%%%%%%%%%%%%%%%%%%%%%%%%%%%%%%%%%%%%%%%%%
\section{Extensions}
\label{sec:extensions}
%%%%%%%%%%%%%%%%%%%%%%%%%%%%%%%%%%%%%%%%%%%%%%%%%%%%%%%%%%%%
%%%%%%%%%%%%%%%%%%%%%%%%%%%%%%%%%%%%%%%%%%%%%%%%%%%%%%%%%%%%
In this section, we outline extensions of our theory to the following settings (cf. figure~\ref{fig:geom}): (i) A nonhomogeneous and isotropic ambient dielectric medium that has a $z$-dependent permittivity~\cite{VolkovMikhailov1988}. (ii) Two semi-infinite coplanar sheets lying in the $xy$-plane, where the EPP propagates along their 
joint boundary, the $y$-axis~\cite{VolkovMikhailov1988}. These sheets have arbitrary and in principle distinct tensor valued surface  conductivities.

%%%%%%%%%%%%%%%%%%%%%%%%%%%%%%%%%%%%%%%%%%%%%%%%%%%%%%%%%%%%%%%
\subsection{Nonhomogeneous ambient medium}
\label{subsec:layered-amb-medium}
%%%%%%%%%%%%%%%%%%%%%%%%%%%%%%%%%%%%%%%%%%%%%%%%%%%%%%%%%%%%%%%
Suppose that the ambient space has a $z$-dependent, bounded dielectric permittivity, $\varepsilon_r(z)$, relative to the vacuum which has permittivity $\varepsilon_0$. For example, for a layered ambient medium
 $\varepsilon_r(z)$ is piecewise constant. In the quasi-electrostatic approximation, define the requisite Green function, $\mathfrak G(x-x',z;z')$, by~\cite{VolkovMikhailov1988}
%%%%%%%%%%%%%%%%%%%%%%%%%%%%%
\begin{equation*}
	\nabla\cdot[\varepsilon_r(z)\nabla \mathfrak G(x,z;z')]=-\delta(x)\,\delta(z-z'),\quad \nabla=(\partial/\partial x, \iu q,\partial/\partial z),
\end{equation*}
%%%%%%%%%%%%%%%%%%%%%%%%%%%%%
with $G(x,z;z')\to 0$ as $\sqrt{x^2+z^2}\to \infty$ for fixed $z'$.
For a semi-infinite sheet, $\Sigma$, which lies in the $xy$-plane with $x>0$, the integral equation for the scalar potential, $\varphi(x)$, results from~(\ref{eq:int-phi-gen}) under the replacement
%%%%%%%%%%%%%%%%%%%%%%%%%
\begin{equation*}
	\mathfrak K(x-x')\rightarrow \mathfrak G(x-x',0;0)\quad\mbox{with}\ k_0=\omega\sqrt{\varepsilon_0\mu}~.
\end{equation*}
%%%%%%%%%%%%%%%%%%%%%%%%%

Let us restrict attention to spatially constant surface conductivities, $\bsig^\Sigma$, given by~(\ref{eq:sigma-matrix}). By following the steps of sections~\ref{sec:solution} and~\ref{sec:EPP_dispersion}, we can derive dispersion relation~(\ref{eq:EPP-dr}) for {\em zero} index $\nu_K$, where $Q_\pm(\xi)$, $\xi^\pm$ and $\mathcal C^\pm$ are given by~(\ref{eq:Q+-})--(\ref{eq:C-coeffs}) with
%%%%%%%%%%%%%%%%%%%%%%%%%
\begin{equation*}
	\mathcal P(\xi)=1+\frac{\iu\omega\mu}{k_0^2}\{\sigma_{xx}\xi^2+(\sigma_{xy}+\sigma_{yx})q\xi+\sigma_{yy}q^2\}\widehat{\mathfrak G}(\xi,0;0)
\end{equation*}
%%%%%%%%%%%%%%%%%%%%%%%%%
and $k_0=\omega\sqrt{\varepsilon_0\mu}$. For {\em nonzero} $\nu_K$, the results follow directly from~\ref{app:nonzero-index}.

For example, suppose $\Sigma$ lies at the interface, at $z=0$, between two dielectric media of constant relative permittivities $\varepsilon_{r1}$ (say, for $z>0$) and $\varepsilon_{r2}$ ($z<0$). Then, 
%%%%%%%%%%%%%%%%%%%%%%%%%%
\begin{equation*}
	\widehat{\mathfrak G}(\xi,0;0)=\frac{1}{\varepsilon_{r1}+\varepsilon_{r2}}(\xi^2+q^2)^{-1/2},\quad \Re\sqrt{\xi^2+q^2}>0.
	\end{equation*}
%%%%%%%%%%%%%%%%%%%%%%%%%%
Accordingly, for $\sigma_{xy}\neq \sigma_{yx}$ the ensuing long-wavelength expansion is [cf.~(\ref{eq:q-longw})]
%%%%%%%%%%%%%%%%%%%%%%
\begin{equation*}
	-\frac{1}{2\pi}\frac{\omega\mu (\sigma_{xy}-\sigma_{yx})}{(\varepsilon_{r1}+\varepsilon_{r2})k_0}\frac{q}{k_0}\Biggl[
	\ln\biggl(\frac{2}{\bq}\biggr)+1 \Biggr]\simeq 1,\ \ \bq=\frac{\iu \omega\mu\sigma_{xx}}{(\varepsilon_{r1}+\varepsilon_{r2})k_0} \frac{q\sg(q)}{k_0}. 
\end{equation*}
%%%%%%%%%%%%%%%%%%%%%%

%%%%%%%%%%%%%%%%%%%%%%%%%%%%%%%%%%%%%%%%%%%%%%%%%%%%%%%%%%%%%%%
%\subsection{Nonlocal surface conductivity}
%\label{subsec:nonlocal-sigma}
%%%%%%%%%%%%%%%%%%%%%%%%%%%%%%%%%%%%%%%%%%%%%%%%%%%%%%%%%%%%%%%

%%%%%%%%%%%%%%%%%%%%%%%%%%%%%%%%%%%%%%%%%%%%%%%%%%%%%%%%%%%%%%%
\subsection{Two coplanar, anisotropic conducting sheets}
\label{subsec:two-sheets}
%%%%%%%%%%%%%%%%%%%%%%%%%%%%%%%%%%%%%%%%%%%%%%%%%%%%%%%%%%%%%%%
 \color{black} Consider two semi-infinite sheets, $\Sigma_L$ and $\Sigma_R$, of distinct conductivities $\bsig^{\Sigma_{L}}$ and $\bsig^{\Sigma_{R}}$. The sheets lie in the $xy$-plane, where $\Sigma_{L}$ ($\Sigma_{R}$) occupies the left (right) half plane with respect to the $y$-axis, for
$x<0$ ($x>0$). The combined material sheet, $\Sigma=\Sigma_L \cup \Sigma_R$, has the surface current density $\bjs(x,y)=e^{\iu q y}\bsig^\Sigma(x)\cdot \bfe_\parallel(x)$, where $\bsig^\Sigma(x)$ is (cf.~\cite{VolkovMikhailov1988})
%%%%%%%%%%%%%%%%%%%%%%%%
\begin{equation*}
	\bsig^\Sigma(x)=\bsig^{\Sigma_L}+ \theta(x) (\bsig^{\Sigma_R} - \bsig^{\Sigma_L})\qquad (\bsig^{\Sigma_L}\neq \bsig^{\Sigma_R},\ -\infty < x < \infty).
\end{equation*}
%%%%%%%%%%%%%%%%%%%%%%%%
Here, $\bsig^{\Sigma_{L,R}}$ are spatially constant $2\times 2$ matrices with positive semidefinite Hermitian parts and elements $\sigma_{\ell \ell'}^{L,R}$ ($\ell, \ell'=x,y$). Suppose that the ambient medium is homogeneous and isotropic with permittivity $\varepsilon$ and permeability $\mu$.

By analogy with~(\ref{eq:phi-int-hom}), the continuous scalar potential $\varphi(x)$, now satisfies
%%%%%%%%%%%%%%%%%%%%%%
\begin{eqnarray*}
\varphi(x)&=& \frac{\iu \omega\mu}{k_0^2} \biggl\{\biggl(\frac{{\rm d}}{\dx},\ \iu q\biggr)\cdot \bsig^{\Sigma_R} \cdot
\left(
\begin{array}{l}
{\rm d}/\dx \cr
\iu q
\end{array}
\right)\int_{0}^\infty \mathfrak K(x-x') \varphi(x')\,\dx'\nonumber\\
&&\mbox +\biggl(\frac{{\rm d}}{\dx},\ \iu q\biggr)\cdot \bsig^{\Sigma_L} \cdot
\left(
\begin{array}{l}
{\rm d}/\dx \cr
\iu q
\end{array}
\right)\int_{-\infty}^0 \mathfrak K(x-x') \varphi(x')\,\dx'\biggr\}\nonumber\\
&& \mbox{} -\frac{\iu\omega\mu}{k_0^2} \biggl(\frac{{\rm d}}{\dx},\ \iu q\biggr)\cdot (\bsig^{\Sigma_R}-\bsig^{\Sigma_L}) \cdot
\left(
\begin{array}{l}
1 \cr
0
\end{array}
\right)
\mathfrak K(x) \varphi_0,\ -\infty< x<\infty,
\end{eqnarray*}
%%%%%%%%%%%%%%%%%%%%%%
with $\varphi_0=\varphi(0)$. Hence, the functional equation for $\hatp_\pm(\xi)$ on the real axis is
%%%%%%%%%%%%%%%%%%%%%%
\begin{equation}\label{eq:func-eq-2sheets}
\mathcal P^L(\xi)\hatp_+(\xi)+\mathcal P^R(\xi)\hatp_-(\xi)=\frac{\omega\mu}{k_0^2} (\sigma_{yx}q+\sigma_{xx}\xi)\hatK(\xi)\varphi_0,
\end{equation}
%%%%%%%%%%%%%%%%%%%%%%
where $\sigma_{\ell \ell'}=\sigma_{\ell \ell'}^R-\sigma_{\ell \ell'}^L$ ($\ell, \ell'=x, y$) and
%%%%%%%%%%%%%%%%%%%%%%
\begin{equation*}
	\mathcal P^{L,R}(\xi)=1+\frac{\iu\omega\mu}{k_0^2} \left\{\sigma_{xx}^{L,R}\xi^2+\left(\sigma_{xy}^{L,R}+\sigma_{yx}^{L,R}\right)q\xi+\sigma_{yy}^{L,R}q^2\right\}\hatK(\xi).
\end{equation*}
%%%%%%%%%%%%%%%%%%%%%%

To carry out the procedures of sections~\ref{sec:solution} and~\ref{sec:EPP_dispersion}, we assume that
%%%%%%%%%%%%%%%%%%%%%%%%%%%%
\begin{equation*}
	\mathcal P^{L,R}(\xi)\neq 0\ \mbox{for\ all\ real}\ \xi.
\end{equation*}
%%%%%%%%%%%%%%%%%%%%%%%%%%%%
In addition, we introduce the suitable index, $\nu_K$, by
%%%%%%%%%%%%%%%%%%%%%%%%%%%%
\begin{equation}\label{eq:index-def-2sheets}
	\nu_K=\frac{1}{2\pi}{\rm arg}\mathcal P(\xi)\biggl|_{\xi=-\infty}^\infty,\quad \mathcal P(\xi)=\frac{\mathcal P^R(\xi)}{\mathcal P^L(\xi)}.
\end{equation}
%%%%%%%%%%%%%%%%%%%%%%%%%%%%
Evidently, $\nu_K=\nu_K^R-\nu_K^L$ where $\nu_K^{R,L}$ is the index associated with $\mathcal P^{R,L}(\xi)$ on the real axis.
In order to derive a dispersion relation for $q$, we take
%%%%%%%%%%%%%%%%%%%%%%%%%%%%
\begin{equation*}
	\nu_K=0.
\end{equation*}
%%%%%%%%%%%%%%%%%%%%%%%%%%%%
Note that the zero value of $\nu_K$ is attained if $\nu_K^R=\nu_K^L$. (In the case with $\bsig^{\Sigma_L}=0$, in the absence of the `left' sheet with $x<0$, we have $\mathcal P^L(\xi)=1$ and thus $\nu_K^L=0$.)

Now introduce the split functions $Q_\pm(\xi)$ via
$Q(\xi)=\ln\mathcal P(\xi)=Q_+(\xi)+Q_-(\xi)$
where $Q_\pm(\xi)$ are given by~(\ref{eq:Q+-}) but $\mathcal P(\xi)=\mathcal P^R(\xi)/\mathcal P^L(\xi)$, as above. Hence, we find
%%%%%%%%%%%%%%%%%%%%%%%%%%%%
\begin{equation*}
e^{-Q_+(\xi)}\hatp_+(\xi)+e^{Q_-(\xi)}\hatp_-(\xi)=\frac{\omega\mu}{k_0^2}
(q\sigma_{yx}+\xi\sigma_{xx})\frac{\hatK(\xi)}{\mathcal P^L(\xi)} e^{-Q_+(\xi)}\varphi_0
\end{equation*}
%%%%%%%%%%%%%%%%%%%%%%%%%%%%
for all real $\xi$. By using the present definition of $Q_\pm(\xi)$, we can expand the right-hand side of the last equation as $-\iu [\Lambda_+(\xi)+\Lambda_-(\xi)]$ where $\Lambda_\pm(\xi)$ are given by~(\ref{eq:Lambda-def}). Note that, in the present setting of two sheets,  $\xi^\pm$ and $\mathcal C^\pm$ are defined by~(\ref{eq:zeros-polyn}) and~(\ref{eq:C-coeffs}) with the substitution $\sigma_{\ell\ell'}=\sigma_{\ell \ell'}^R-\sigma_{\ell \ell'}^L$ ($\ell, \ell'=x,y$). Accordingly, invoking~\ref{app:Q-asympt} with $Q(\xi)=Q^R(\xi)-Q^L(\xi)$ and $Q^{R,L}(\xi)=\ln\mathcal P^{R,L}(\xi)$, we obtain~(\ref{eqs:phi-invFT}) for $\varphi(x)$.

After some manipulation of the Fourier integrals for $\varphi(x)$, we find the limits
%%%%%%%%%%%%%%%%%%%%%%%%%%%%%%%%%%%%%%%%%%%%%%%%%%%
\begin{eqnarray}\label{eq:varphi-limits-2sheets}
	\varphi(0^{\pm})&=&\varphi_0\mp \varphi_0 \mathcal A(q,\omega)\left\{\frac{1}{2\pi\iu}\lim_{\epsilon\downarrow 0}\int_{-\infty}^\infty \frac{\xi\, e^{\mp Q_\mp(\xi)}}{(\xi-\xi^+)(\xi-\xi^-)}e^{\pm \iu\xi\epsilon}\,\dxi \right. \nonumber\\
	 && \  \left. -\frac{\xi^\mp}{\xi^+-\xi^-} e^{\mp Q_\mp (\xi^\mp)}\right\},\quad \varphi(0^\pm)=\lim_{x\downarrow 0\atop x\uparrow 0}\varphi(x),
\end{eqnarray}
%%%%%%%%%%%%%%%%%%%%%%%%%%%%%%%%%%%%%%%%%%%%%%%%%%%
where
%%%%%%%%%%%%%%%%%%%%%%%%%%%%%%%%%%%%%%%%%%%%%%%%%%%
\begin{equation*}
\mathcal A(q,\omega)=\mathcal C^+ e^{-Q_+(\xi^+)}+\mathcal C^- e^{Q_-(\xi^-)}.
\end{equation*}
%%%%%%%%%%%%%%%%%%%%%%%%%%%%%%%%%%%%%%%%%%%%%%%%%%%
These limits are consistent with the continuity of $\varphi(x)$ across the edge if $\mathcal A(q,\omega)=0$, which yields an equation of form~(\ref{eq:EPP-dr}) for the EPP dispersion relation. The respective long-wavelength expansion follows from~(\ref{eq:q-longw}) with $\sigma_{\ell\ell'}=\sigma_{\ell \ell'}^R-\sigma_{\ell \ell'}^L$ (cf. \cite{Stauber2019}). \color{black}

The theory for the present setting of two sheets can be extended to include a nonhomogeneous ambient medium as outlined in section~\ref{subsec:layered-amb-medium}.
For {\em nonzero} index $\nu_K$, the results for two sheets are analogous to those for a single sheet (see~\ref{app:nonzero-index}).

%%%%%%%%%%%%%%%%%%%%%%%%%%%%%%%%%%%%%%%%%%%%%%%%%%%%%%%%%%%%%%%
%%%%%%%%%%%%%%%%%%%%%%%%%%%%%%%%%%%%%%%%%%%%%%%%%%%%%%%%%%%%%%%
\section{Discussion and conclusion}
\label{sec:conclusion}
%%%%%%%%%%%%%%%%%%%%%%%%%%%%%%%%%%%%%%%%%%%%%%%%%%%%%%%%%%%%%%%
%%%%%%%%%%%%%%%%%%%%%%%%%%%%%%%%%%%%%%%%%%%%%%%%%%%%%%%%%%%%%%%

In this paper, we derived the dispersion relation for the EPP that propagates along the edge of a semi-infinite flat conducting sheet of arbitrary tensor valued and spatially homogeneous conductivity in a uniform medium in the nonretarded frequency regime. Our findings have the following ingredients. (i) We produce the quasi-electrostatic approximation via the separation of two scales. (ii) We connect the existence of the EPP dispersion relation to the topological notion of the {\em index}, $\nu_K$, for Wiener-Hopf integral equations~\cite{Krein1962}.  (iii) We derive an asymptotic formula for the EPP wave number at long wavelengths which manifests a generalized effect of anisotropy. Our analysis applies to inhomogeneous ambient media and pairs of coplanar sheets.

The index $\nu_K$ is an integer that depends on the material, frequency, and EPP wave number, $q$, and quantifies how the EPP direction of propagation affects the number of all, exponentially decaying and growing in the direction normal to the sheet, types of nonretarded bulk wave modes. \color{black} For the EPP dispersion relation to exist, $\nu_K$ must vanish. For many physical situations, this condition is expected to hold if the  number of bulk wave modes propagating away from the edge remains invariant under the reversal of the EPP direction of propagation. In contrast, $\nu_K$ may possibly be nonzero if the sum of the off-diagonal elements of $\bsig^\Sigma$ does not vanish. If $\nu_K>0$ then the EPP wave number takes continuum values for every frequency compatible with this $\nu_K$. In contrast, if $\nu_K<0$ no EPP can exist in the respective region of $(q,\omega)$. %Such a transition of the index value in principle corresponds to the crossover of a branch of the magnetoplasmon energy found in~\cite{VolkovMikhailov1988} in the appropriate limit of vanishing dissipation and anisotropy. This correspondence is further developed and discussed in a separate paper~\cite{MSLLM-preprint}. 

Our results inspire exciting open questions. For example, one should explore the possibility for topologically protected edge plasmons on semi-infinite conducting sheets, in the purely anisotropic and hyperbolic regimes~\cite{Nemilentsau2016}. The problem with a conducting nanoribbon sheet, which may exhibit standing waves because of multiple reflections from the ends, deserves some attention because of its close experimental relevance. We considered a model with a local tensor surface conductivity. Models of nonlocal surface conductivities such as the ones emerging from viscous hydrodynamic or kinetic models of the 2D electron system are the subjects of work in progress.

 \color{black}

\section*{Acknowledgements}
The authors (DM, MM, TS, TL and ML) thank Alex Levchenko for bringing reference \cite{CohenGoldstein2018} to their attention. DM, MM and ML acknowledge partial support by the ARO MURI award W911NF-14-1-0247, and the Institute for Mathematics and its Applications (NSF grant DMS-1440471) at the University of Minnesota for several visits. The research of DM was also partially supported by the NSF under grant DMS-1517162, and by a Research and Scholarship award by the Graduate School, University of Maryland in the spring of 2019. MM also acknowledges partial support by the NSF under grant DMS-1912847. The work of TS has been supported by Spain's MINECO under grant   
FIS2017-82260-P as well as by the CSIC Research Platform on Quantum  
Technologies PTI-001. TL acknowledges support from the NSF through the University of Minnesota MRSEC under grant DMR-1420013, and the grant NSF/EFRI-1741660. ML was also supported by the NSF under grant DMS-1906129.

\appendix

%%%%%%%%%%%%%%%%%%%%%%%%%%%%%%%%%%%%%%%%%%%%%%%%%%%%%%%%%%%%%%%%%%%%
\section{On the evaluation of the index $\nu_K$}
\label{app:index-eval}
%%%%%%%%%%%%%%%%%%%%%%%%%%%%%%%%%%%%%%%%%%%%%%%%%%%%%%%%%%%%%%%%%%%%

In this appendix, we provide a plausibility argument that yields~(\ref{eq:index-value}) for the index $\nu_K=\nu_K(q)$ associated with the function $\mathcal P(\xi;q)$ on the real axis. 

Recall that $\mathcal P(\xi;q)$ is defined by~(\ref{eq:P-def}). We introduce the `dual' function
%%%%%%%%%%%%%%%%%%%%%%%%%%%%%%%%%%%%%%%%%%%%%%%%%%%%%%%%%%%%%%
\begin{equation}
	\mathcal P_*(\xi;q)=1+\frac{1}{k_{\rm sp}}\frac{(\xi-\xi^+)(\xi-\xi^-)}{\sqrt{\xi^2+q^2}},\quad \Re\sqrt{\xi^2+q^2}>0,
\end{equation}
%%%%%%%%%%%%%%%%%%%%%%%%%%%%%%%%%%%%%%%%%%%%%%%%%%%%%%%%%%%%%%
where $k_{\rm sp}=\iu 2k_0^2/(\omega\mu\sigma_{xx})$. This $\mathcal P_*(\xi;q)$ amounts to waves that grow exponentially with $|z|$. Note that $\mathcal P_*(\xi;q)$ comes from $\mathcal P(\xi;q)$  with the replacement $\sqrt{\xi^2+q^2}\rightarrow -\sqrt{\xi^2+q^2}$. In this vein, the `dual' index $\nu_{K*}(q)$ for $\mathcal P_*(\xi;q)$ on the real axis is 
%%%%%%%%%%%%%%%%%%%%%%%%%%%%%%%%%%%%%%%%%%%%%%%%%%%%%%%%%%%%%%
\begin{equation}\label{eq:index-adjoint-def}
\nu_{K*}(q)=\frac{1}{2\pi}\arg\mathcal P_*(\xi;q)\bigl|_{\xi=-\infty}^{+\infty}.	
\end{equation}
%%%%%%%%%%%%%%%%%%%%%%%%%%%%%%%%%%%%%%%%%%%%%%%%%%%%%%%%%%%%%%
For ease of notation, we drop the subscript `$K$' in $\nu_K$ and $\nu_{K*}$. Consider the relation
%%%%%%%%%%%%%%%%%%%%%%%%%%%%%%%%%%%%%%%%%%%%%%%%%%%%%%%%%%%%
\begin{equation*}
	-k_{\rm sp}^2(\xi^2+q^2)\mathcal P(\xi) \mathcal P_*(\xi)=(\xi-\xi^+)(\xi-\xi^-)-k_{\rm sp}^2(\xi^2+q^2). 
\end{equation*}
%%%%%%%%%%%%%%%%%%%%%%%%%%%%%%%%%%%%%%%%%%%%%%%%%%%%%%%%%%%%
Hence, the roots of $\mathcal P(\xi;q)=0$ in the top Riemann sheet coincide with the zeros $\xi$ of the quartic polynomial on the right-hand side that satisfy $\Re\{(\xi-\xi^+)(\xi-\xi^-)/k_{\rm sp}\}>0$.

It is advantageous to consider the quantity
%%%%%%%%%%%%%%%%%%%%%%%%%%%%%%%%%%%%%%%%%%%%%%%%%%%%%%%%%%%%%%
\begin{equation*}
\nu(q)-\nu(-q)=\frac{1}{2\pi \iu}
\int_{-\infty}^\infty \left\{\frac{\mathcal P'(\xi;q)}{\mathcal P(\xi;q)}+\frac{\mathcal P'(-\xi;q)}{\mathcal P(-\xi;q)}\right\}\,\dxi=2\nu(q),	
\end{equation*}
%%%%%%%%%%%%%%%%%%%%%%%%%%%%%%%%%%%%%%%%%%%%%%%%%%%%%%%%%%%%%%
by use of the identity $\nu(-q)=-\nu(q)$. Here, the prime denotes differentiation with respect to $\xi$.
By deforming the integration path in the upper half $\xi$-plane, we find
%%%%%%%%%%%%%%%%%%%%%%%%%%%%%%%%%%%%%%%%%%%%%%%%%%%%%%%%%%%%%%
\begin{equation}\label{eq:nu-diff}
2\nu(q)=N^+(q) -N^-(q) +I_{\rm bc}(q),
\end{equation}
%%%%%%%%%%%%%%%%%%%%%%%%%%%%%%%%%%%%%%%%%%%%%%%%%%%%%%%%%%%%%%
where $N^{\pm}(q)$ is the number of zeros of $\mathcal P(\xi;q)$ in the upper ($+$) or lower ($-$) half $\xi$-plane in the top Riemann sheet, and
%%%%%%%%%%%%%%%%%%%%%%%%%%%%%%%%%%%%%%%%%%%%%%%%%%%%%%%%%%%%%%
\begin{equation*}
I_{\rm bc}(q)=\frac{1}{2}\frac{1}{2\pi\iu}\sum_{\varsigma=\pm}\int_{\Gamma_{\rm bc}}\left\{\frac{\mathcal P'(\varsigma\xi;q)}{\mathcal P(\varsigma\xi;q)}-\frac{\mathcal P_*'(\varsigma\xi;q)}{\mathcal P_*(\varsigma\xi;q)}\right\}\, \dxi.
\end{equation*}
%%%%%%%%%%%%%%%%%%%%%%%%%%%%%%%%%%%%%%%%%%%%%%%%%%%%%%%%%%%%%%
The (oriented) path $\Gamma_{\rm bc}$ is wrapped in the counterclockwise sense around the branch cut, which emanates from $\xi=\iu q{\rm sg}(q)$, for $\sqrt{\xi^2+q^2}$ in the upper half $\xi$-plane. 

In a similar vein, we can relate $\nu_*(q)$ with the integral $I_{\rm bc}(q)$. We obtain 
%%%%%%%%%%%%%%%%%%%%%%%%%%%%%%%%%%%%%%%%%%%%%%%%%%%%%%%%%%%%%%
\begin{equation}\label{eq:nu-star-diff}
2\nu_*(q)=N_*^+(q) -N_*^-(q) -I_{\rm bc}(q),
\end{equation}
%%%%%%%%%%%%%%%%%%%%%%%%%%%%%%%%%%%%%%%%%%%%%%%%%%%%%%%%%%%%%%
where $N_*^{\pm}(q)$ is the number of zeros of $\mathcal P_*(\xi;q)$ in the upper ($+$) or lower ($-$) half $\xi$-plane in the first Riemann sheet, where $\Re\sqrt{\xi^2+q^2}>0$.  Therefore, by eliminating $I_{\rm bc}(q)$ from~(\ref{eq:nu-diff}) and~(\ref{eq:nu-star-diff}), we have
%%%%%%%%%%%%%%%%%%%%%%%%%%%%%%%%%%%%%%%%%%%%%%%%%%%%%%%%%%%%%%
\begin{equation}\label{eq:nu-nu*}
\nu(q)+\nu_*(q)=\frac{N^+(q) - N^-(q)}{2}+\frac{N_*^+(q) -N_*^-(q)}{2}.
\end{equation}
%%%%%%%%%%%%%%%%%%%%%%%%%%%%%%%%%%%%%%%%%%%%%%%%%%%%%%%%%%%%%%
The right-hand side of this formula only takes the values $0, \pm 1, \pm 2$. This is evident by $N_*^+ +N_*^- +N^+ + N^-=4$, which entails $\nu(q)+\nu_*(q)=N^+ + N_*^+ -2$.
Since we have not found an explicit formula for $I_{\rm bc}(q)$ in terms of $N^{\pm}$ or $N_*^{\pm}$, we resort to an empirical claim about the resulting value of $\nu(q)$. For an appreciable range of physical parameters which we used in numerics, we obtain 
$\nu_*(q)=0$.
Hence, (\ref{eq:nu-nu*}) 
yields~(\ref{eq:index-value}). 

We close this appendix with two numerical examples for 
$\nu(q)$ to corroborate~(\ref{eq:index-value}). With the setting and notation of remark~5 (section~\ref{sec:EPP_dispersion}),  we consider the following cases: 

\noindent (a) $\bar{\boldsymbol{\sigma}}^\Sigma(0)={\rm diag}(0.001+0.1 \iu, 0.002+0.2\iu)$, $\phi=0.4\pi$ and $\bar q=0.850 \bar q_a$ where $\bar q_a$ solves~(\ref{eq:EPP-dr}) under the conductivity $\bar{\boldsymbol{\sigma}}^\Sigma(\phi)$ (case (iii) in remark 5). The resulting $\mathcal P(\xi;q)$ has 2 zeros in the lower half $\xi$-plane of the first Riemann sheet, with $N^+=0$ and $N^-=2$, while $N_*^+=1=N_*^-$. By~(\ref{eq:index-def}) we find $\nu_K=-1$, in agreement with~(\ref{eq:index-value}). By setting $\bar q=-0.850 \bar q_a$, we verify by~(\ref{eq:index-def}) that $\nu_K$ changes sign thus becoming $\nu_K=1$.

\noindent (b) $\bar{\boldsymbol{\sigma}}^\Sigma(0)={\rm diag}(0.001+0.1 \iu, 0.002+0.2\iu)$ and $\phi=0.166\pi$. First, we solve~(\ref{eq:EPP-dr}) and obtain $\bar q\simeq 16.438 + 0.164\iu=\bar q_b$, while we find $\nu_K=0$ by definition~(\ref{eq:index-def}). Then, we scale down this $\bar q$ by taking $\bar q= 0.75 \bar q_b$, thus abandoning relation~(\ref{eq:EPP-dr}). The ensuing $\mathcal P(\xi;q)$ has 3 zeros in the first Riemann sheet with $N^+=0$ and $N^-=3$, and only 1 zero in the second Riemann sheet with $N_*^+=1$ (and $N_*^-=0$). By~(\ref{eq:index-def}) we directly find $\nu_K=-1$, in agreement with formula~(\ref{eq:index-value}).

%%%%%%%%%%%%%%%%%%%%%%%%%%%%%%%%%%%%%%%%%%%%%%%%%%%%%%%%%%%%%%%%%%%%
\section{On the asymptotics of $Q_\pm(\xi)$ for large $|\xi|$}
\label{app:Q-asympt}
%%%%%%%%%%%%%%%%%%%%%%%%%%%%%%%%%%%%%%%%%%%%%%%%%%%%%%%%%%%%%%%%%%%%
In this appendix, we sketch the derivation of asymptotic formulas for the split functions $Q_\pm(\xi)$, which are defined by~(\ref{eq:Q+-}), as $\xi\to \infty$ (cf.~\cite{MML2017}).

Let us focus on $Q_+(\xi)$ for $\Im\,\xi>0$. By the change of variable $\chi=\xi'/\xi$, we write
%%%%%%%%%%%%%%%%%%%%%
\begin{equation*}
Q_+(\xi)=\frac{1}{2\pi \iu}	\int_0^{\infty e^{-\iu \arg\xi}}\frac{\chi[Q(\xi\chi)-Q(-\xi\chi)]+[Q(\xi\chi)+Q(-\xi\chi)]}{\chi^2-1}\,\dch.
\end{equation*}
%%%%%%%%%%%%%%%%%%%%%
By considering $\chi$ as fixed, we write
%%%%%%%%%%%%%%%%%%%%%
\begin{equation*}
	Q(\xi\chi)-Q(-\xi\chi)=\ln\left\{\frac{1+\frac{\iu\omega\mu\sigma_{xx}\xi}{2k_0^2} \frac{(\chi-\xi^+/\xi)(\chi-\xi^-/\xi)}{\sqrt{\chi^2+q^2/\xi^2}}}{1+\frac{\iu\omega\mu\sigma_{xx}\xi}{2k_0^2} \frac{(\chi+\xi^+/\xi)(\chi+\xi^-/\xi)}{\sqrt{\chi^2+q^2/\xi^2}}}\right\}=Q_1(\xi\chi),
\end{equation*}
%%%%%%%%%%%%%%%%%%%%%
noting that this $Q_1(\xi\chi)$ is $\mathcal O(\xi^{-1})$ as $\xi\to\infty$. In a similar vein, we have
%%%%%%%%%%%%%%%%%%%%%%%%
\begin{eqnarray*}
	\lefteqn{Q(\xi\chi)+Q(-\xi\chi)=2\ln\biggl(\frac{\iu \omega\mu\sigma_{xx}\xi}{2k_0^2}\,\chi\biggr)+Q_2(\xi\chi),}\nonumber\\
	Q_2(\xi\chi)&=&\mathcal O(1/\xi)\ \mbox{as}\ \xi\to\infty.
\end{eqnarray*}
%%%%%%%%%%%%%%%%%%%%%%%

By use of the above formulas in the integral for $Q_+(\xi)$, we directly find
%%%%%%%%%%%%%%%%%%%%%%%%%%%%%%%%%%%%%%%%%%%%%
\begin{equation}\label{eq:Q+_asympt}
Q_+(\xi)= \frac{1}{2} \ln\biggl(\frac{\omega\mu\sigma_{xx}\xi}{2k_0^2}\biggr)+\mathcal O\biggl(\frac{\ln\xi+1}{\xi}\biggr)\ \mbox{as}\ \xi\to\infty\quad (\Im\ \xi>0).
\end{equation}
%%%%%%%%%%%%%%%%%%%%%%%%%%%%%%%%%%%%%%%%%%%%%
By reflection through the real axis in the $\xi$-plane, we assert that, for $\Im\,\xi<0$,
%%%%%%%%%%%%%%%%%%%%%%%%%%%%%%%%%%%%%%%%%%%%%
\begin{equation}\label{eq:Q-_asympt}
Q_-(\xi)= \frac{1}{2} \ln\biggl(-\frac{\omega\mu\sigma_{xx}\xi}{2k_0^2}\biggr)+\mathcal O\biggl(\frac{\ln\xi+1}{\xi}\biggr)\ \mbox{as}\ \xi\to\infty.
\end{equation}
%%%%%%%%%%%%%%%%%%%%%%%%%%%%%%%%%%%%%%%%%%%%%

%%%%%%%%%%%%%%%%%%%%%%%%%%%%%%%%%%%%%%%%%%%%%%%%%%%%%%%%%%%%%%%%%%%%
\section{On the spectral theory for nonzero index $\nu_K$}
\label{app:nonzero-index}
%%%%%%%%%%%%%%%%%%%%%%%%%%%%%%%%%%%%%%%%%%%%%%%%%%%%%%%%%%%%%%%%%%%%
In this appendix, we formally present the mathematical theory for the nonzero index $\nu_K$, defined by~(\ref{eq:index-def}), that is associated with $\mathcal P(\xi)$ on the real axis (see also~\cite{Krein1962}).

\medskip

{\em Negative index, $\nu_K<0$.} We will show that it is impossible to find $q$ such that~(\ref{eq:phi-int-hom}) admits nontrivial integrable and continuous solutions. The starting point is functional equation~(\ref{eq:func-eq-xi}). The first task is to construct a function, $\Phi(\xi)\mathcal P(\xi)$, with {\em zero index} which replaces $\mathcal P(\xi)$. Subsequently, we follow the steps of section~\ref{sec:solution} for zero index.

Accordingly, define the function~\cite{Krein1962,Masujima-book}
%%%%%%%%%%%%%%%%%%%%%%
\begin{equation}\label{eq:Phi-def-neg}
	\Phi(\xi)=\prod_{j=1}^{|\nu_K|}\Biggl(\frac{\xi-z_{j}^+}{\xi-p_{j}^-}\Biggr)
\end{equation}
%%%%%%%%%%%%%%%%%%%%%%
where $\{z_{j}^+\}_{j=1}^{|\nu_K|}$ and $\{p_{j}^-\}_{j=1}^{|\nu_K|}$ consist of arbitrary points in the upper ($+$) and lower ($-$) half planes. Notice that the index related to $\Phi(\xi)$ on $\mathbb{R}$ is equal to $|\nu_K|$, and $\Phi(\xi)\to 1$ as $\xi\to\infty$. The former property implies that
%%%%%%%%%%%%%%%%%%%%%%
\begin{equation*}
	\Phi(\xi)\mathcal P(\xi)\ \mbox{has\ zero\ index\ on\ the\ real\ axis}.
\end{equation*}
%%%%%%%%%%%%%%%%%%%%%%
By analogy with section~\ref{sec:solution}, we write $\Phi(\xi)\mathcal P(\xi)=e^{Q_+(\xi)} e^{Q_-(\xi)}$ and then find suitable functions $Y_\pm(\xi)$ such that
$\mathcal P(\xi)=Y_-(\xi)/Y_+(\xi)$. Thus, (\ref{eq:func-eq-xi}) becomes
%%%%%%%%%%%%%%%%%%%%%%
\begin{equation}\label{eq:func-eq-mod}
Y_+(\xi)\hatp_+(\xi)+Y_-(\xi)\hatp_-(\xi)=\frac{\omega\mu}{2k_0^2}
\frac{q\sigma_{yx}+\xi\sigma_{xx}}{\sqrt{\xi^2+q^2}}Y_+(\xi)\varphi_0
\end{equation}
%%%%%%%%%%%%%%%%%%%%%%
for $\xi$ in the real axis. By virtue of the Cauchy integral formula, we assert that
%%%%%%%%%%%%%%%%%%%%%%
\begin{equation}\label{eq:Q-mod-def}
	Q_\pm(\xi)=\pm \frac{1}{2\pi\iu}\int_{-\infty}^\infty \frac{\ln[\Phi(\xi')\mathcal P(\xi')]}{\xi'-\xi} \ \dxi'\quad (\pm\Im\,\xi>0).
\end{equation}
%%%%%%%%%%%%%%%%%%%%%
Furthermore, without loss of generality, we can take
%%%%%%%%%%%%%%%%%%%%%
\begin{equation*}
	Y_-(\xi)=e^{Q_-(\xi)}\Biggl(\prod_{j=1}^{|\nu_K|}(\xi-z_j^+)\Biggr)^{-1},\
	Y_+(\xi)=e^{-Q_+(\xi)}\Biggl(\prod_{j=1}^{|\nu_K|}(\xi-p_j^-)\Biggr)^{-1}.
\end{equation*}
%%%%%%%%%%%%%%%%%%%%%
Alternate choices for $Y_\pm(\xi)$ are possible but yield the same final conclusion after lengthier algebra.
In view of~\ref{app:Q-asympt}, we have $Q_\pm(\xi)=\frac{1}{2}\ln\xi+O(1)$ as $\xi\to\infty$.

Next, we determine suitable split functions $\Lambda_\pm(\xi)$ such that
%%%%%%%%%%%%%%%%%%%%%%
\begin{equation*}
\frac{\omega\mu}{2k_0^2}
\frac{q\sigma_{yx}+\xi\sigma_{xx}}{\sqrt{\xi^2+q^2}}Y_+(\xi)\varphi_0=-i[\Lambda_+(\xi)+\Lambda_-(\xi)].	
\end{equation*}
%%%%%%%%%%%%%%%%%%%%%%
A close inspection of the left-hand side by use of the definitions of $Q_\pm(\xi)$ entails
%%%%%%%%%%%%%%%%%%%%%%
\begin{eqnarray}
	\Lambda_+(\xi)&=&\frac{\mathcal C^-}{\xi-\xi^-}\bigl[Y_-(\xi^-)-Y_+(\xi)\bigr]
	-\frac{\mathcal C^+}{\xi-\xi^+} \bigl[Y_+(\xi)-Y_+(\xi^+)\bigr],\nonumber\\
	\Lambda_-(\xi)&=&\frac{\mathcal C^+}{\xi-\xi^+}\bigl[Y_-(\xi)-Y_+(\xi^+)\bigr]+\frac{\mathcal C^-}{\xi-\xi^-}\bigl[Y_-(\xi)-Y_-(\xi^-)\bigr],\label{eq:Qcase-def}
\end{eqnarray}
%%%%%%%%%%%%%%%%%%%%%%
where $\xi^\pm$ and $\mathcal C^\pm$ are given by~(\ref{eq:zeros-polyn}) and~(\ref{eq:C-coeffs}).

Without further ado, from the functional equation for $\hatp_\pm(\xi)$ and the asymptotic behavior of $\Lambda_\pm(\xi)$ and $Q_\pm(\xi)$ for large
$|\xi|$, we derive the formulas
%%%%%%%%%%%%%%%%%%%%%%%%%%
\begin{equation}\label{eq:hatphi+-_neg}
\hatp_-(\xi)=-\iu \Lambda_-(\xi) Y_-(\xi)^{-1}\varphi_0,\
	\hatp_+(\xi)=-\iu \Lambda_+(\xi) Y_+(\xi)^{-1} \varphi_0.
\end{equation}
%%%%%%%%%%%%%%%%%%%%%%%%%%
By using the formulas for $\Lambda_\pm(\xi)$ and $Y_\pm(\xi)$ we see that $\Lambda_+(\xi)Y_+(\xi)^{-1}$ contains $\mathcal O(\xi^{|\nu_K|-1/2})$ and $\mathcal O(\xi^{|\nu_K|-3/2})$ terms as $\xi\to\infty$. In an effort to satisfy the limits $\hatp_+(\xi)\to 0$ as $\xi\to \infty$ and $\varphi(x)\to \varphi_0$ as $x\uparrow 0$, we need to eliminate such terms.  We realize that it is impossible to do so unless $\varphi_0=0$ (which yields the trivial solution).

\medskip

{\em Positive index, $\nu_K> 0$.} We show that, for any given $\omega$, (\ref{eq:phi-int-hom}) may in principle admit nontrivial integrable and continuous solutions if $q$ lies in some region of the complex plane that is compatible with the prescribed $\nu_K$. We start with the construction of a function, $\Phi(\xi)\mathcal P(\xi)$, of zero index and then describe $\varphi(x)$ in the spirit of section~\ref{sec:solution}.

Define the function~\cite{Krein1962,Masujima-book}
%%%%%%%%%%%%%%%%%%%%%%
\begin{equation}\label{eq:Phi-def-pos}
	\Phi(\xi)=\prod_{j=1}^{\nu_K}\Biggl(\frac{\xi-z_{j}^-}{\xi-p_{j}^+}\Biggr)
\end{equation}
%%%%%%%%%%%%%%%%%%%%%%
where $\{z_{j}^-\}_{j=1}^{\nu_K}$ and $\{p_{j}^+\}_{j=1}^{\nu_K}$ are formed by arbitrary points in the lower ($-$) and upper ($+$) half planes. Hence, the index related to $\Phi(\xi)$ on $\mathbb{R}$ is equal to $-\nu_K$, which means that $\Phi(\xi) \mathcal P(\xi)$ has zero index on $\mathbb{R}$. By writing $\Phi(\xi)\mathcal P(\xi)=e^{Q_+(\xi)} e^{Q_-(\xi)}$, where $Q_\pm(\xi)$ are given by~(\ref{eq:Q-mod-def}), we need to find $Y_\pm(\xi)$ such that $\mathcal P(\xi)=Y_-(\xi)/Y_+(\xi)$. Without loss of generality, minimizing the algebra without affecting the final conclusion, we set
%%%%%%%%%%%%%%%%%%%%%
\begin{equation*}
	Y_-(\xi)=e^{Q_-(\xi)}\prod_{j=1}^{\nu_K}(\xi-p_j^+),\
	Y_+(\xi)=e^{-Q_+(\xi)}\prod_{j=1}^{\nu_K}(\xi-z_j^-).
\end{equation*}
%%%%%%%%%%%%%%%%%%%%%
Functional equation~(\ref{eq:func-eq-xi}) then becomes of form~(\ref{eq:func-eq-mod}) with the above choices of $Y_\pm(\xi)$.

Next, we determine split functions $\Lambda_\pm(\xi)$ for the right-hand side of~(\ref{eq:func-eq-mod}). Thus, formulas~(\ref{eq:Qcase-def}) are recovered for the present case. By our choice for $Y_\pm(\xi)$, we have
%%%%%%%%%%%%%%%%%%%%%
\begin{equation*}
	\Lambda_+(\xi)=\mathcal O(\xi^{\nu_K-3/2}),\ \Lambda_-(\xi)=\mathcal O(\xi^{\nu_K-1/2})\ \mbox{as}\ \xi\to\infty.
\end{equation*}
%%%%%%%%%%%%%%%%%%%%%
The rearrangement of terms in the functional equation for $\hatp_\pm(\xi)$ gives
%%%%%%%%%%%%%%%%%%%%%
\begin{equation*}
	Y_+(\xi)\hatp_+(\xi)+\iu \Lambda_+(\xi)\varphi_0=-Y_-(\xi)\hatp_-(\xi)-\iu \Lambda_-(\xi)\varphi_0\quad (-\infty<\xi<\infty).
\end{equation*}
%%%%%%%%%%%%%%%%%%%%%

By analytic continuation to complex $\xi$, we set the function of each side of the last equation equal to an entire function, $\mathfrak E(\xi)$. Taking into account the asymptotic behaviors of $\Lambda_\pm(\xi)$ and $Y_\pm(\xi)$ as $\xi\to\infty$, and the integrability of $\varphi(x)$, we see that
%%%%%%%%%%%%%%%%%%%%%
\begin{equation*}
	\mathfrak E(\xi)\ \mbox{cannot\ grow\ as\ fast\ as}\ \xi^{\nu_K-1/2}\ \mbox{in\ the\ limit}\ \xi\to\infty.
\end{equation*}
%%%%%%%%%%%%%%%%%%%%%
By Liouville's theorem, we assert that $\mathfrak E(\xi)$ is an $(\nu_K-1)$-th degree polynomial, viz.,
%%%%%%%%%%%%%%%%%%%%%
\begin{equation*}
	\mathfrak E(\xi)=\sum_{j=0}^{\nu_K-1}c_j \xi^j
\end{equation*}
%%%%%%%%%%%%%%%%%%%%%
where the coefficients, $c_j$, are viewed as arbitrary complex numbers. Thus, we obtain
%%%%%%%%%%%%%%%%%%%%%
\begin{equation}\label{eq:hatphi+-_pos}
	\hatp_\mp(\xi)=-\iu \Lambda_\mp(\xi) Y_\mp(\xi)^{-1}\varphi_0
	\mp Y_\mp(\xi)^{-1}\mathfrak E(\xi).
\end{equation}
%%%%%%%%%%%%%%%%%%%%%

The Fourier inversion of these formulas implies that $\varphi(x)$ is expressed as a linear superposition of suitable, independent functions with coefficients $c_j$ ($j=0, \ldots, \nu_K-1$). We set $c_{\nu_K-1}=0$ to eliminate an unwanted divergence for $\varphi(x)$ as $x\uparrow 0$. Hence, $\mathfrak E(\xi)$ is a polynomial of degree $\nu_K-2$ ($\mathfrak E= 0$ identically if $\nu_K=1$). In fact, we find
%%%%%%%%%%%%%%%%%%
\begin{equation*}
\lim_{x\downarrow 0\atop (x\uparrow 0)}\int_{-\infty}^\infty e^{\iu \xi x} Y_\mp(\xi)^{-1}\mathfrak E(\xi)\, \dxi=
\int_{-\infty}^\infty Y_\mp(\xi)^{-1}\mathfrak E(\xi)\, \dxi=0,
\end{equation*}
%%%%%%%%%%%%%%%%%%
by closing the integration path in the lower ($-$) or upper ($+$) half plane where $Y_\mp(\xi)^{-1}$ is analytic.
Thus, the second term on the right-hand side of~(\ref{eq:hatphi+-_pos}) does not affect $\varphi(0)$.

It remains to show that $\lim_{x\uparrow 0}\varphi(x)=\varphi_0=\lim_{x\downarrow 0}\varphi(x)$ without any further condition on $q$. After some algebra for $\lim_{x\uparrow 0}\varphi(x)$, we encounter the expression
%%%%%%%%%%%%%%%%%%
\begin{eqnarray*}
\lefteqn{\lim_{x\uparrow 0}\varphi(x)= \varphi_0 \frac{\iu\omega\mu\sigma_{xx}}{2k_0^2}\big[\mathcal C^+ Y_+(\xi^+)+\mathcal C^- Y_-(\xi^-)\big]\frac{1}{2\pi\iu} \int_{-\infty}^\infty \frac{Y_-(\xi)^{-1}}{\sqrt{\xi^2+q^2}}\xi\,\dxi}\nonumber\\
&& \mbox{} -\varphi_0 \frac{\iu\omega\mu\sigma_{xx}}{2k_0^2} \big[\mathcal C^+Y_+(\xi^+)\xi^-+\mathcal C^-Y_-(\xi^-)\xi^+\big]\frac{1}{2\pi \iu} \int_{-\infty}^\infty \frac{Y_-(\xi)^{-1}}{\sqrt{\xi^2+q^2}}\,\dxi.
\end{eqnarray*}
%%%%%%%%%%%%%%%%%%
By use of the factorization $\mathcal P(\xi)=Y_-(\xi)/Y_+(\xi)$, we find
%%%%%%%%%%%%%%%%%%
\begin{equation*}
\frac{\iu\omega\mu\sigma_{xx}}{2k_0^2}\frac{Y_-(\xi)^{-1}}{\sqrt{\xi^2+q^2}}=\frac{Y_+(\xi)^{-1}-Y_-(\xi)^{-1}}{(\xi-\xi^+)(\xi-\xi^-)}.
\end{equation*}
%%%%%%%%%%%%%%%%%%
Hence, by application of the residue theorem, the limit in question is equal to
%%%%%%%%%%%%%%%%%%%%
\begin{eqnarray*}
\lefteqn{\lim_{x\uparrow 0}\varphi(x)= \varphi_0 \big[\mathcal C^+ Y_+(\xi^+)+\mathcal C^- Y_-(\xi^-)\big] \left[\xi^+ \frac{Y_+(\xi^+)^{-1}}{\xi^+-\xi^-}+\xi^-\frac{Y_-(\xi^-)^{-1}}{\xi^--\xi^+}\right]} \nonumber\\
&&\mbox{} -\varphi_0 \big[\mathcal C^+ Y_+(\xi^+)\xi^-+\mathcal C^- Y_-(\xi^-)\xi^+\big]
\left[\frac{Y_+(\xi^+)^{-1}}{\xi^+-\xi^-}+\frac{Y_-(\xi^-)^{-1}}{\xi^--\xi^+}\right]\nonumber\\
&=& \varphi_0 (\mathcal C^+ + \mathcal C^-)=\varphi_0,
\end{eqnarray*}
%%%%%%%%%%%%%%%%%%%%
for {\em any} complex $q$ allowed by the given value of $\nu_K$. Similarly, one can show that $\lim_{x\downarrow 0}\varphi(x)=\varphi_0$, with no restriction on $q$. We leave the proof to the reader.

%%%%%%%%%%%%%%%%%%%%%%%%%%%%%%%%%%%%%%%%%%%%%%%%%%%%%%%%%%%%%%%%%%%%
\section{Long-wavelength expansion via Mellin transform}
\label{app:MT}
%%%%%%%%%%%%%%%%%%%%%%%%%%%%%%%%%%%%%%%%%%%%%%%%%%%%%%%%%%%%%%%%%%%%
In this appendix, we outline a framework for the expansion of dispersion relation~(\ref{eq:EPP-dr})
when $|q|$ is sufficiently small, and derive~(\ref{eq:Q-exp-loworder}). To this end, we formally apply the Mellin transform technique, which is amenable to systematic calculations~\cite{SasielaShelton1993,Fikioris-asympt-book}.

Recalling that $\bq= i\omega\mu\sigma_{xx} q\sg(q)/(2k_0^2)$ with $\Re\,q\neq 0$, define
%%%%%%%%%%%%%%%%%%%%%%
\begin{eqnarray*}
	f^\pm(\bq)&=&\pm 2\pi \iu Q_\pm(\xi^\pm)\sg(q)=\int_0^\infty
	 \frac{\dzet}{\zeta^2-(\alpha^\pm)^2} \biggl\{\zeta\bigl[\ln(1+\bq A(\zeta))\nonumber\\
	 && \mbox{} -\ln(1+\bq A(-\zeta))\bigr]
	+\alpha^\pm  \bigl[\ln(1+\bq A(\zeta))+\ln(1+\bq A(-\zeta))\bigr]\biggr\}.
\end{eqnarray*}
%%%%%%%%%%%%%%%%%%%%%%
Note that the mapping $q\to -q$ implies $f^\pm\to f^\mp$ whereas $\breve q$ is unchanged. Let us take $\Re\, q>0$. For later algebraic convenience, consider $\bq >0$. We analytically continue the results to complex $\bq$ in the end. The Mellin transform of $f^\pm(\bq)$ is defined by
%%%%%%%%%%%%%%%%%%%%%%
\begin{equation}\label{eq:MT-def}
\widetilde f^{\pm}(s)=\int_0^\infty \bq^{-s} f^\pm(\bq)\,\dq,\quad \alpha_1 <\Re\, s<\alpha_2,
\end{equation}
%%%%%%%%%%%%%%%%%%%%%%
where $\alpha_{1,2}$ are such that the integral converges. The inverse Mellin transform is
%%%%%%%%%%%%%%%%%%%%
\begin{equation}\label{eq:invMT-def}
f^\pm(\bq)=\frac{1}{2\pi \iu}\int_{\gamma-\iu \infty}^{\gamma+\iu \infty} \widetilde f^\pm(s) \bq^{s-1}\,\ds\quad (\alpha_1 <\gamma <\alpha_2).
\end{equation}
%%%%%%%%%%%%%%%%%%%%
This technique works if, for example, the asymptotic expansion of $f^\pm(\bq)$ as $\bq\to 0$ consists of terms of the form $\bq^{\ell} (\ln\bq)^m$ for positive integers $m$ and $\ell$. These terms can be extracted from~(\ref{eq:invMT-def}) via the shift of the integration path to the right and collection of residues of $\widetilde f^\pm(s)\breve q^{s-1}$ at poles located in $\{s\in\mathbb{C}:\,\Re\,s\ge \alpha_2\}$.

We carry out this task to low orders in $\bq$ by use of the
gamma function, $\Gamma(w)$~\cite{Bateman-I}. After some algebra, we obtain
%%%%%%%%%%%%%%%%%%%%%%
\begin{eqnarray}\label{eq:tildef-calc}
\widetilde f^\pm(s)&=&\frac{\Gamma(2-s) \Gamma(s-1)}{s-1}\int_0^\infty  \frac{\dzet}{\zeta^2-(\alpha^\pm)^2}\biggl\{\zeta \big[A(\zeta)^{s-1}-A(-\zeta)^{s-1}\big]\nonumber\\
&& \mbox{} +\alpha^\pm \big[A(\zeta)^{s-1}+A(-\zeta)^{s-1}\big]\biggr\},\ 1<\Re\, s<2.
\end{eqnarray}
%%%%%%%%%%%%%%%%%%%%%%
This $\widetilde f^\pm(s)$ is meromorphic, with poles at $s=n\in\mathbb{Z}$ in the $s$-plane. For our purpose of deriving an expansion for $f^\pm(\bq)$ if $|\bq|$ is small, only the poles of $\widetilde f^\pm(s)$ at $s=2, 3,\ldots$ are relevant. By expanding $\widetilde f^\pm(s)$ around the double pole at $s=2$, we have
%%%%%%%%%%%%%%%%%%%%%%%%%%%
\begin{eqnarray*}
\widetilde f^\pm(s)&\sim& 2 \frac{1+\epsilon}{\epsilon}\biggl\{\big[-\alpha^\mp+(\alpha^++\alpha^-)\epsilon]\int_0^\infty \frac{\zeta^{1-\epsilon}}{\zeta^2-(\alpha^{\pm})^2}\,\dzet \nonumber\\
 && \mbox{} +\alpha^{\mp} \int_0^\infty \biggl(\frac{\zeta}{\zeta^2-(\alpha^\pm)^2}-\frac{1}{\sqrt{1+\zeta^2}}\biggr)\,\dzet\biggr\},\ \epsilon=2-s\to 0,
\end{eqnarray*}
%%%%%%%%%%%%%%%%%%%%%%%%%%%
where $\Re(\alpha^\pm)^2 < 0$. By evaluating the requisite integrals for $\Re\,\epsilon>0$~\cite{Bateman-I}, we obtain
%%%%%%%%%%%%%%%%%%%%%%%%%%%%%%
%\begin{equation*}
%\widetilde f^\pm(s)\sim 2\left\{ -\frac{\alpha^\mp}{\epsilon^2}+\big[\alpha^+ +\alpha^- -\alpha^\mp -\alpha^{\mp}\ln\,2\big] \frac{1}{\epsilon}\right\},
%\end{equation*}
%%%%%%%%%%%%%%%%%%%%%%%%%%%%%%
%%%%%%%%%%%%%%%%%%%%%%%
\begin{equation*}
\widetilde f^\pm(s) \bq^{s-1}\sim 2\bq \biggl\{-\frac{\alpha^\mp}{\epsilon^2}-\biggl[\alpha^\mp \ln\biggl(\frac{2}{\bq}\biggr)-\alpha^\pm\biggr]\frac{1}{\epsilon}\biggr\},\ \epsilon=2-s\to 0.
\end{equation*}
%%%%%%%%%%%%%%%%%%%%%%%
By shifting the integration path and evaluating the residue at $s=2$ in~(\ref{eq:invMT-def}), we find
%%%%%%%%%%%%%%%%%%%%%%
\begin{equation}\label{eq:f-exp-q}
f^\pm(\bq)\sim -2 \bq \biggl\{ \alpha^\mp \ln\biggl(\frac{2}{\bq}\biggr)-\alpha^\pm\biggr\}\ \mbox{if}\ |\breve q|\ll 1, \ \Re\,q>0,
\end{equation}
%%%%%%%%%%%%%%%%%%%%%%
where any correction terms, not shown here, are due to the residues from poles of $\widetilde f^\pm(s)$ at $s=3, 4,\ldots$. If $\Re\,q<0$, the asymptotic formula  for $f^\pm(\bq)$ results from~(\ref{eq:f-exp-q}) by switching the $+$ and $-$ superscripts on the right-hand side. These formulas can be analytically continued to complex values of $\bq$ via properties of the logarithm.

Expansion~(\ref{eq:Q-exp-loworder}) follows from~(\ref{eq:f-exp-q}) and its counterpart for $\Re\,q<0$. To compute higher orders in $\breve q$, one can extract residues for $f^\pm(\bq)$ as $s$ approaches poles at $3, 4, \ldots$.

%%%%%%%%%%%%%%%%%%%%%%%%%%%%%%%%%
%%%%%%%%%%%%%%%%%%%%%%%%%%%%%%%%%

\end{document}